\documentclass[12pt]{article}
\usepackage{graphicx}
\usepackage{enumerate}
\usepackage{natbib}
\usepackage{url} 

\newcommand{\blind}{0}

\usepackage{amsfonts, amsmath, mathrsfs, bm}
\usepackage{float, multirow, booktabs, threeparttable}
\usepackage{array}
\usepackage{tikz}
\usepackage[figuresright]{rotating}

\usepackage{xcolor}

\usepackage{color}

\newcommand{\PreserveBackslash}[1]{\let\temp=\\#1\let\\=\temp}
\newcolumntype{C}[1]{>{\PreserveBackslash\centering}p{#1}}
\newcolumntype{R}[1]{>{\PreserveBackslash\raggedleft}p{#1}}
\newcolumntype{L}[1]{>{\PreserveBackslash\raggedright}p{#1}}

\newtheorem{theorem}{Theorem}

\newtheorem{example}{Example}
\newtheorem{assumption}{Assumption}

\newcommand{\ipw}{ {\rm IPW} }
\newcommand{\elw}{ {\rm ELW} }
\newcommand{\mse}{ {\mathbb{M}\rm se}}
\newcommand{\tr}{ \mbox{\bf tr}}

\newcommand{\convergeto}{\overset{d}{\longrightarrow}}
\def\T{{ \mathrm{\scriptscriptstyle \top} }}
\newcommand{\bas}{\begin{eqnarray*}}
\newcommand{\eas}{\end{eqnarray*}}

\newcommand{\ba}{\begin{eqnarray}}
\newcommand{\ea}{\end{eqnarray}}

\newcommand{\bit}{\begin{itemize}}
\newcommand{\eit}{\end{itemize}}

\newcommand{\ben}{\begin{enumerate}}
\newcommand{\een}{\end{enumerate}}

\newcommand{\e}{ { \mathbb{E}}}
\newcommand{\var}{ { \mathbb{V} {\rm ar}}}

\usepackage{hyperref}

\begin{document}

\def\spacingset#1{\renewcommand{\baselinestretch}%
{#1}\small\normalsize} \spacingset{1}


\if0\blind
{
\title{
Nearly optimal capture--recapture sampling and
empirical likelihood weighting estimation
for M-estimation with big data
}

\vspace{-0.1in}
\date{}

\author{Yan Fan, Yang Liu, Yukun Liu, and Jing Qin
\footnote{
Yan Fan is an associate professor at the
School of Statistics and Information, Shanghai University of
International Business and Economics, Shanghai, 
China (Email: \emph{fanyan@suibe.edu.cn}).
Yang Liu is a postdoctoral researcher at
KLATASDS-MOE,
School of Statistics,
East China Normal University,
Shanghai, 
China
(Email: \emph{liuyangecnu@163.com}).
Yukun Liu is a professor at
KLATASDS-MOE,
School of Statistics,
East China Normal University,
Shanghai, 
China (Email: \emph{ykliu@sfs.ecnu.edu.cn}).
Jing Qin is a mathematical statistician at the
National Institute of Allergy and Infectious Diseases,
National Institutes of Health, Frederick, Maryland, USA (Email: \emph{jingqin@niaid.nih.gov}).
This research is supported by the National Key R\&D Program of China
(2021YFA1000100 and 2021YFA1000101),
the National Natural Science Foundation
of China (11971300, 12101239, 12171157, 71931004),
the Natural Science Foundation of Shanghai (19ZR1420900),
the China Postdoctoral Science Foundation (Grant 2020M681220), and
the 111 Project (B14019).
Dr. Yukun Liu is the corresponding author.
The first two authors contributed equally to this paper.
}
}
\maketitle
} \fi

\if1\blind
{
\bigskip
\begin{center} \Large \bf
Nearly optimal capture--recapture sampling and
empirical likelihood weighting estimation
for M-estimation with big data
\end{center}
\medskip
} \fi

\begin{abstract}
Subsampling techniques can reduce the computational costs of processing big data.
Practical subsampling plans typically involve initial
uniform sampling and refined sampling.
With a subsample, big data inferences are generally built on
the inverse probability weighting (IPW),
which becomes unstable when the probability weights are close to zero
and cannot incorporate auxiliary information.
First, we consider capture--recapture sampling, which combines
an initial uniform sampling with a second Poisson sampling.
Under this sampling plan,
we propose an empirical likelihood weighting (ELW)
estimation approach to an M-estimation parameter.
Second, based on the ELW method,
we construct a nearly optimal capture--recapture sampling plan
that balances estimation efficiency and computational costs.
Third, we derive methods for determining the smallest sample sizes
with which the proposed sampling-and-estimation method
produces estimators of guaranteed precision.
Our ELW method overcomes the instability of IPW
by circumventing the use of inverse probabilities,
and utilizes auxiliary information
including the size and certain sample moments of big data.
We show that the proposed ELW method
produces more efficient estimators than IPW,
leading to more efficient optimal sampling plans
and more economical sample sizes for a prespecified estimation precision.
These advantages are
confirmed through simulation studies and real data analyses.
\end{abstract}

\noindent%
{\it Keywords}:
Big data; Capture--recapture sampling;
Empirical likelihood;
M-estimation; 
Sample size formula.
\vfill

\newpage
\spacingset{1.9} 

\section{Introduction}
\label{sec:intro}

One of the most significant features of big data is its incredibly large volume,
which poses serious challenges to its timely processing.
Data analytics need to be performed efficiently
so that the results are made available to users
in a cost-effective and timely manner.
A popular and efficient strategy for solving this problem
is to draw small-scale subsamples from the big data (original sample)
and make statistical inferences based on the subsamples
\citep{drineas2006sampling, drineas2011faster}.
Compared with the original big data,
the subsamples are usually much smaller, and
so subsample-based inferences significantly reduce the required computational resources.

Subsample-based inferences for big data
generally involve two fundamental issues:
how to draw an effective subsample and
how to make efficient statistical inferences
based on the subsample.
Regarding the first issue, it is generally accepted that
carefully designed sampling probabilities
make unequal probability samplings more efficient than
simple random or uniform sampling.
Many researchers have developed efficient or optimal sampling plans for
frequently encountered parametric statistical problems,
including linear regression models  
\citep{ma2014statistical},
logistic regression  
\citep{fithian2014local, wang2019more},
generalized linear models  
\citep{ai2019optimal},
quantile regression  
\citep{ai2021optimal, fan2021optimal, wang2021optimal},
and more general models \citep{shen2021surprise, yu2020optimal}. 
All of the aforementioned methods use sampling with replacement,
except those of 
\cite{fithian2014local}, \cite{wang2019more}, \cite{shen2021surprise},
and \cite{yu2020optimal},
which consider Poisson sampling, one of the easiest-to-implement sampling without replacement systems.
Under Poisson sampling, the samples are independently drawn
according to Bernoulli experiments with prespecified
success probabilities.
Poisson sampling has two advantages over sampling with replacement:
it never draws replicate observations
and its implementation is
free from memory constraints
\citep{yao2019optimal}.

For the second issue, subsample-based statistical inferences
for big data are usually performed through inverse probability weighting (IPW),
which leads to the Hansen--Hurwitz estimator
\citep{hansen1943theory}
under sampling with replacement
and to the Horvitz--Thompson estimator
\citep{horvitz1952generalization}
under sampling without replacement.
However, the subsample-based
IPW estimation procedure for big data analysis suffers from
two weaknesses.
First, the IPW estimator can be highly unstable
if there are extremely small probabilities,
resulting in poor finite-sample performance of
the accompanying asymptotic-normality-based inferences
\citep{kang2007demystifying, robins2007comment,
cao2009improving, imbens2009recent, busso2014new, han2019general}.
This weakness of the IPW estimator
has been observed in survey sampling \citep{zong2019improved}
and areas such as missing data problems
\citep{robins2007comment}, treatment effect estimation 
\citep{crump2009dealing, khan2010irregular, yang2018asymptotic},
and survival analysis 
\citep{robins2000correcting, dong2020inverse}.
To circumvent this notorious issue,
an unnatural lower boundedness assumption is often imposed on the probabilities
\citep{rosenbaum1983central, mccaffrey2013inverse, sun2018inverse}. However,
tiny probabilities are frequently encountered in practice,
especially when the propensity scores
are estimated from data
\citep{yang2018asymptotic,ma2020robust}.
Second, the efficiency of IPW cannot be enhanced
by incorporating auxiliary information,
although this is often available in big data analysis.
For example, the sample mean of some variables
in a big dataset can be quickly calculated
at little computational cost;
this can be taken as auxiliary information when
inferences are made based on a subsample.
To overcome the first limitation of IPW,
\cite{liu2021biased} proposed a biased-sample empirical likelihood (EL) weighting
method to serve the same general purpose as IPW,
which completely overcomes the instability of IPW-type estimators
by circumventing the use of inverse probabilities.
However, their EL method does not take into account
the auxiliary information defined by general estimating equations.

In the case of big data, the optimal sampling depends on the statistical
problem under study and the accompanying subsample-based estimation procedure.
To consider both the generality and convenience
of theoretical analysis and implementation,
we focus on M-estimation problems with convex loss functions,
and consider the use of Poisson sampling.
Popular examples of M-estimation problems with convex loss functions
include linear regression, quantile regression,
and many generalized linear regressions (e.g.,
logistic regression, softmax regression, and Poisson regression). 
The sampling probabilities of the ideal optimal samplings
depend on the ideal parameter estimator from the big data itself.
For the optimal sampling to be practically applicable,
an initial sample is required to produce
an initial estimate of the parameter of interest.
In this paper, we regard each of two samplings as a capture,
and hence regard the whole sampling procedure as a
capture--recapture sampling. This is a novel viewpoint
in the study of subsampling for big data.
Capture--recapture sampling is widely used
to estimate population sizes in biology, ecology,
and reliability studies 
\citep{mcCrea2014analysis}.
A significant difference
between capture--recapture
sampling for big data analysis
and the equivalent methods for biology, ecology,
and reliability studies is
that the ``population size'' is known in the former,
whereas it is unknown, and constitutes the target parameter to be estimated,
in the latter.

This paper makes three contributions to
the literature of subsample-based big data analysis.
\ben
\item
First, we develop an empirical likelihood weighting (ELW) estimation method for a capture--recapture sample
from big data, incorporating auxiliary information defined by estimating equations.
The proposed estimation procedure
not only overcomes the instability of the IPW
by circumventing the use of inverse probabilities,
but also achieves enhanced efficiency by incorporating auxiliary information.
We show that, in theory, the proposed ELW estimator is asymptotically
more efficient than the IPW estimator.

\item
Second, balancing the estimation efficiency with the computational costs,
we construct a nearly optimal capture--recapture sampling plan
by minimizing the upper bound of
the asymptotic mean square error (MSE) of the proposed ELW estimator.
The sample from the first capture is used to estimate the subsampling probabilities
of the second capture. 

\item
Third, we determine the minimal sample size needed
so that the proposed nearly optimal sampling plan
achieves the desired precision requirement
in terms of MSE and absolute error.
As the ELW estimator is more efficient than
the IPW estimator,
the proposed nearly optimal capture--recapture sampling
is expected to outperform
existing optimal IPW-based subsampling plans.
\een

The remainder of this paper is organized as follows.
In Section \ref{sec:ELW}, after introducing
the M-estimation problem
and the commonly used IPW estimation method,
we introduce the ELW estimation procedure
with auxiliary information under
a general capture--recapture sampling plan,
and study the asymptotic behavior of the ELW estimator.
In Section \ref{sec:samplan}, we construct a nearly optimal
capture--recapture sampling plan and discuss
its practical implementation.
In Section \ref{sec:sampsize}, we derive
the minimal sample size needed for
the proposed estimator to meet a prespecified precision.
Simulation studies and real applications
are reported in Sections \ref{sec:simu} and \ref{sec:data}.
Finally, Section \ref{sec:disc} concludes with a discussion.
All technical proofs are given in the supplementary material for clarity.

\section{Empirical likelihood weighting estimation}\label{sec:ELW}

\subsection{Setup and IPW}

Suppose that the big data
consist of $N$ observations $\{Z_i\}_{i=1}^N$, which
are independent and identically distributed (i.i.d.) copies
from a population $Z$ with an unknown cumulative distribution function $F$.
Parametric models indexed by a $q$-dimensional parameter $\theta$
are usually imposed to extract information from data.
Let $\ell(z, \theta)$ be a user-specific convex loss function
that quantifies the lack-of-fit of a parametric model indexed
by a parameter $\theta$ based on an observation $z$.
The average loss or risk function is
\(
R(\theta) = \e \{ \ell(Z, \theta)\} = \int \ell(z, \theta)dF(z)
\).
We define the parameter of interest $\theta_0$ to be
the risk minimizer 
\citep{huber2004robust, shen2021surprise}
\ba
\label{target}
\theta_0 = \arg\min_{\theta} R(\theta).
\ea
This setup includes many common problems as special cases.
When $Z$ is a scalar,
the true parameter value $\theta_0$ is the mean or median of $Z$
if $\ell(Z, \theta) = (Z-\theta)^2$ or $|Z-\theta|$.
When $Z=(Y, X^\T)^\T$, $\theta_0$ may be the
population-level regression coefficient
in the generalized linear regression, least-squares regression,
quantile regression, and
expectile regression models under the specification of $\ell(z; \theta)$
given in Table \ref{tab:example}.

\begin{table}
\centering 
\setlength{\columnsep}{.05in}
\caption{
Loss functions  and the matrix $V$ under commonly-used regression models.
  Here   $\ddot a(x)$ is the
second derivative  of $a(x)$.
}
\label{tab:example}
\begin{threeparttable}
\begin{tabular}{ccc}
\toprule
Regression model  & $\ell(z; \theta)$   & $V$\\
\midrule
Generalized linear  &  $- yx^\T\theta + a(x^\T\theta) - \log\{b(y)\}$
&   $\e \left[ XX^\T \ddot{ a}(X^\T \theta_0 \right]$  \\
Poisson   & $-yx^\T\theta + \exp(x^\T\theta) + \log(y!)$
&    $\e \{ XX^\T \exp(X^\T \theta_0) \}$ \\
Logistic  & $-yx^\T\theta + \log\{1 + \exp(x^\T\theta)\}$
&   $\e \left[ XX^\T \frac{\exp(x^\T \theta_0)}{\{1 + \exp(x^\T  \theta_0)\}^2} \right]$ \\
Least square  & $(y - x^\T\theta)^2$ &   $\e( XX^\T )$  \\
Quantile   & $(y - x^\T\theta)\{ \tau - I(y - x^\T\theta < 0)\}$
&  $\e\{ XX^\T f(X^\T\theta_0 \mid X)\}$ \\
Expectile  &  $(y - x^\T\theta)^2 |\tau - I(y - x^\T\theta < 0)|$
&   $\e\{ XX^\T |\tau - I(Y \leq X^\T \theta_0 )|\}$ \\
\bottomrule
\end{tabular}
\end{threeparttable}
\end{table}

Based on the big-data observations,
$\hat \theta_N = \arg\min_{\theta} \sum_{i=1}^N \ell(Z_i, \theta) $
is the ideal estimator of $ \theta $.
For massive datasets, $N$ can be
so large that the direct calculation of $\hat \theta_N$ is formidable or practically infeasible.
Subsampling techniques then come into play to reduce the computation costs.
As discussed in the introduction,
we consider the use of capture--recapture sampling,
where the first capture is a Poisson sampling with
an equal sampling probability
and the second capture is another Poisson sampling, but with
generally unequal sampling probabilities.
Let the unequal sampling probabilities in the second capture be
$\pi_i= \pi(Z_i)$, $i=1, \ldots, N$, for some function $\pi(\cdot)$.
The ideal sample sizes for both the Poisson samplings
in the capture--recapture sampling plan,
$r_0$ and $r = \sum_{i=1}^N \pi_i$, must be specified beforehand.

In the first capture,
for each $Z_i$ ($i=1, 2, \ldots, N$),
we conduct a Bernoulli experiment with success probability $\alpha_{10} = r_0 /N$
and denote the result as $D_{i1}$,
which is equal to 1 for success and 0 otherwise.
Datum $Z_i$ is sampled in the first capture
if and only if $D_{i1}=1$.
The samples in the first capture are used to produce
an initial estimate of $\theta$,
which is them employed to determine the sampling probabilities of the second capture.
For now, we assume that the $\pi_i$ are known.
In the second capture,
we again conduct a Bernoulli experiment,
but with success probability $\pi_i$
for datum $Z_i$, and denote the result as $D_{i2}$;
in the second capture, datum $Z_i$ is sampled
if and only if $D_{i2}=1$.
Finally, the resulting capture--recapture sample
can be written as
$\{ (D_iZ_i, D_{i1}, D_{i2}), \; i=1, 2, \ldots, N \}$,
where $D_{i} = I(D_{i1}+D_{i2}>0)$ and $I(\cdot)$ is the indicator function.

\begin{assumption}
\label{assumption-data}
The $N$ random vectors $(Z_{i}, D_{i1}, D_{i2})$ ($i=1, \ldots, N$)
are i.i.d. copies of $(Z, D_{(1)}, D_{(2)})$.
Suppose that the distribution $F(z)$ of $Z$ is nondegenerate,
$\e(D_{(1)}|Z ) = \e(D_{(1)}) = \alpha_{10}$,
$\e(D_{(2)}|Z) = \pi(Z)$,
and $\alpha_{20} = \e(D_{(2)} ) = \e\{ \pi(Z) \}$. 
\end{assumption}

Let $D=I(D_{(1)}+ D_{(2)}>0)$, where $D_{(1)}$ and $D_{(2)}$
are as defined in Assumption \ref{assumption-data}.
Then, $\e(D) = 1 - \{1 - \e(D_1)\}\{1 - \e(D_2)\} =
1 - (1 - \alpha_{10})(1 - \alpha_{20})$.
For a given datum $Z$,
the overall probability of being sampled is
$\varphi(Z ) = \e(D \mid Z) = 1- (1-\alpha_{10})\{ 1-\pi(Z)\}$ under Assumption \ref{assumption-data}.
Based on the capture--recapture sample,
the IPW estimator of $\theta$ is
\ba
\label{ipw}
\hat\theta_{\ipw} = \arg\min_{\theta} \hat R_{\ipw}(\theta)
\equiv \arg\min_{\theta} \frac{1}{N} \sum_{i=1}^N \frac{D_i}{\varphi(Z_i) } \ell (Z_i, \theta),
\ea
where
\(
\hat R_{\ipw}(\theta)
\)
is the IPW estimator of the risk function
$
R(\theta)
$.

\begin{assumption}
\label{assumption-loss}
Suppose that $\ell(z, \theta)$ is a loss function that is
convex with respect to $\theta$,
and that
$
\ell (z, \theta_0+t) = \ell (z, \theta_0) +
\dot{\ell}(z)^\T t + \xi(z, t)
$ holds in a neighborhood of $t= 0$.
Here,
$\dot{\ell}(z) = \partial \ell(z, \theta_0)/\partial \theta $
satisfies $\e\{\dot{\ell}(Z)\} = 0$ and
$B_{ \dot{\ell} \dot{\ell}} = \e \{ \dot{\ell}(Z) \dot{\ell}^\T(Z)/ \varphi(Z ) \}$
is finite,
and $\xi(z, t)$ satisfies
$
\e \{ \xi(Z, t) \} = (1/2) t^\T V t+ o(\| t\|^2)
$
and
$\e\{ \xi^2(Z, t) \} = o(\| t\|^2)$ 
for a positive-definite matrix $V$
as
$\| t \| \rightarrow 0$.

\end{assumption}
Assumption \ref{assumption-loss} is satisfied by many common
regression models, such as those in Table \ref{tab:example},
where the corresponding matrice $V$ are also provided
for convenience of applications.

\begin{theorem}
\label{asym-ipw}
Suppose that Assumptions \ref{assumption-data} and \ref{assumption-loss} are satisfied
and that $ \alpha_{10}, \alpha_{20} \in (0,1)$ are fixed quantities.
As $N$ goes to infinity,
$\sqrt{N}(\hat \theta_{\ipw} - \theta_0) \convergeto \mathcal{N}(0, \Sigma_{\ipw})$,
where $\convergeto$ denotes
``converges in distribution to'' and
$\Sigma_{\ipw} = V^{-1}B_{ \dot{\ell} \dot{\ell}}V^{-1}.$
\end{theorem}

As discussed in the introduction,
if some probabilities $\varphi(Z_i)$ are too close to zero,
$\hat R_{\ipw}(\theta)$ exhibits remarkable instability,
making the resulting IPW estimator $\hat \theta_\ipw$ in \eqref{ipw}
undesirably unstable.
In the context of big data analysis, auxiliary information
is often available. For example, the response mean of a big data sample
can often be quickly calculated with little extra effort,
and can be regarded as auxiliary information in subsample-based analysis.
However, the estimation efficiency of the IPW method cannot be enhanced by incorporating auxiliary information.
Based on optimal estimating function theory \citep{godambe1960},
the score function derived from the complete-data likelihood is optimal
in the class of inverse weighting estimating functions
\citep[Section 5.2]{qin2017}.
This motivates us to consider the full-likelihood-based inference
approach under the capture--recapture sampling.

\subsection{ELW estimation
under capture--recapture sampling}

Given the capture--recapture data $\{ (D_iZ_i, D_{i1}, D_{i2}), \; i=1, 2, \ldots, N \}$,
the full likelihood is
\ba
\label{likelihood}
{N \choose n} \prod_{i=1}^N \left[
\{ \varphi(Z_i) dF(Z_i) \}^{D_i} \cdot (1-\alpha)^{1-D_i}\right],
\ea
where $\alpha = \e(D) = \int \varphi(z) dF(z) $
is the marginal probability of observing a value of $Z$.
The true value of $\alpha$ is $\alpha_0 = 1-(1-\alpha_{10})(1-\alpha_{20})$
under Assumption \ref{assumption-data}.
Following \cite{liu2021biased}, we use
the empirical likelihood method \citep{owen1988empirical, owen2001empirical} to handle
$F(z)$. Using the principle of the empirical likelihood,
we model $F(z)$ by a step function
\(
\sum_{i=1}^N p_i I(Z_i\leq z)
\), where the $p_i$ are positive and sum to one.

Then, the full log-likelihood becomes the empirical log-likelihood
\ba
\label{log-el}
\sum_{i=1}^N [ D_i \log(p_i)
+ D_i \log\{ \varphi(Z_i) \} + (1-D_i)\log (1-\alpha ) ],
\ea
where
the feasible $p_i$ satisfy
$p_i\geq 0$,
$\sum_{i=1}^N p_i=1$,
and
\ba
\label{constr-pi1}
\sum_{i=1}^N p_i\{\varphi(Z_i) - \alpha \}=0.
\ea
The previous equation follows from $\alpha = \int \varphi(z) dF(z)$.
The $Z_i$ with $D_i=0$ are not observed.
Although appearing in the expression of the above likelihood,
they do not actually contribute to the likelihood.
The expression of the empirical log-likelihood implies that only those
$p_i$ with $D_i=1$ make a contribution to the likelihood.

If we take $\alpha$ to be an unknown parameter,
\cite{liu2021biased} showed that the maximum point of \eqref{log-el}
under the constraints $p_i\geq 0$,
$\sum_{i=1}^N p_i=1$, and \eqref{constr-pi1} is always well defined
if there are at least two different values in
$\{\varphi(Z_i): D_i=1, i=1,2, \ldots, N\}$ (or, equivalently, $\{\pi(Z_i): D_i=1, i=1,2, \ldots, N\}$).
\cite{liu2021biased} took the resulting $p_i$, say $\tilde p_i$, as the weights
and proposed a biased-sample empirical likelihood weighting estimation
method that serves the same purpose as IPW, but overcomes the problem of instability.
Regardless of whether it is known or not,
the parameter $\alpha $ is treated as
an unknown quantity in their method. As a result, their weighting method is always well defined,
as their focus was to develop a new weighting method
that is insensitive to small inclusion probabilities.

Under the two Poisson samplings
in the capture--recapture sampling,
the true parameter values $\alpha_{10}$ and $\alpha_{20}$
need to be prespecified prior to their implementation,
so that $\alpha_0 = 1-(1-\alpha_{10})(1-\alpha_{20})$ is known a priori.
Unlike \cite{liu2021biased},
we make full use of this and other auxiliary information to improve the efficiency of
the resulting point estimator of $\theta$.
The feasible $p_i$ should satisfy
\ba
\label{constr-pi2}
\sum_{i=1}^N p_i\{\varphi(Z_i) - \alpha_0 \}=0.
\ea
In addition, for massive datasets, although
solving the optimization problem $ \min \sum_{i=1}^N \ell(Z_i, \theta)$
is complicated and time-consuming,
the big data sample mean $\sum_{i=1}^N Z_i/N $
or other sample moments can be calculated relatively easily.
This can be taken as auxiliary information
when we make statistical inferences about the big data
based on a subsample.
Suppose that $\bar h = (1/N)\sum_{i=1}^N h(Z_i)$ is available
for some function $h $, which may be vector-valued.
For convenience, we assume that $\e \{ h(Z)\} = 0$
is known. In practice, we recommend replacing $h(Z)$ by $h(Z) - \bar h$.
This can be formulated as one more estimating equation:
\ba
\label{constr-h}
\sum_{i=1}^N p_i h (Z_i) = 0.
\ea

In summary, we recommend estimating the $p_i$ by
their maximum empirical likelihood estimator,
which is the maximizer of the empirical log-likelihood \eqref{log-el}
under the constraints $p_i\geq 0$,
$\sum_{i=1}^N p_i=1$ and \eqref{constr-pi2}, \eqref{constr-h}. 
No nondegenerate solution to this optimization problem exists
if the constraints do not hold simultaneously \citep{chen2008adjusted, liu2010adjusted},
or, equivalently, if the origin lies outside of
the convex hull of $\{ h_e(Z_i): D_i=1, 1\leq i \leq N \}$,
where $h_e(Z) = (\varphi(Z) - \alpha_0, \ h^\T (Z) )^\T$.
In this situation, the optimal weights $ \hat p_i$ are undefined, and
we define them to be $\tilde p_i$, which are the maximizers of
\eqref{log-el} under the constraints $p_i\geq 0$,
$\sum_{i=1}^N p_i=1$ and \eqref{constr-pi1}
in the case of unknown $\alpha$; see \cite{liu2021biased}.
Otherwise, by the Lagrange multiplier method, we have
\ba
\label{pi-opt}
\hat p_i = \frac{1}{\sum_{j=1}^N D_j} \cdot \frac{D_i}{1+ \hat \lambda^\T h_e(Z_i)},
\ea
where $ \hat \lambda $ is the solution to
\(
\sum_{i=1}^N D_i h_e(Z_i)/\{ 1+\hat \lambda^\T h_e(Z_i) \} = 0.
\)

Given $\hat p_i$,
we propose to estimate $\theta$
by the ELW
estimator
\ba
\label{ELW}
\hat \theta_{\elw} = \arg\min_{\theta} \hat R_{\elw }(\theta)\equiv
\arg\min_{\theta} \sum_{i=1}^N \hat p_i \ell(Z_i, \theta)
\ea
where $\hat R_{\elw }(\theta)$
is the ELW estimator of the risk function
$
R(\theta)
$.
If the loss function $\ell(z, \theta)$ is differentiable with respect to $\theta$
for almost all $z$,
an alternative ELW estimator of $\theta$
can be obtained by
maximizing the empirical log-likelihood \eqref{log-el} under the constraints
$p_i\geq 0$,
$\sum_{i=1}^N p_i=1$ with \eqref{constr-pi2}, \eqref{constr-h},
and $\sum_{i=1}^N p_i \partial \ell(Z_i, \theta)/\partial \theta =0$.
Because the dimensions of $\theta$
and $\partial \ell(Z_i, \theta)/\partial \theta$ are the same,
the resulting maximum EL estimator
is exactly equal to $\hat \theta_{\elw} $.

\begin{theorem}
\label{asym-el}
Suppose that Assumptions \ref{assumption-data} and \ref{assumption-loss} hold,
$B_{ hh} = \e \{ h_e(Z)h_e^\T(Z)/ \varphi(Z ) \}$ is positive-definite, and
$ \alpha_{10}, \alpha_{20} \in (0,1)$ are fixed and known.
As $N$ goes to infinity,
\ben
\item[(a)]
$\hat \theta_{\elw}$
is consistent with $\theta_0$ and
$
\sqrt{N}(\hat \theta_{\elw} - \theta_0 )
=
-V^{-1} \cdot N^{1/2} \sum_{i=1}^N \hat p_i \dot{\ell}(Z_i) + o_p(1)
$;
\item[(b)]
$\sqrt{N}(\hat \theta_{\elw} - \theta_0 )
\convergeto \mathcal{N}(0, \Sigma_{\elw})$
with $
\Sigma_{\elw}
=
V^{-1} ( B_{ \dot{\ell} \dot{\ell}} - B_{ \dot{\ell} h} B_{hh}^{-1} B_{ \dot{\ell} h}^\T )V^{-1}
$, where $B_{ \dot{\ell} h} = \e \{ \dot{\ell} (Z)h_e^\T(Z)/ \varphi(Z ) \}$ and
$B_{ \dot{\ell} \dot{\ell}} = \e \{ \dot{\ell}(Z) \dot{\ell}^\T(Z)/ \varphi(Z ) \}$;
\item[(c)]
If the auxiliary information defined by \eqref{constr-h} is ignored,
then $\sqrt{N}(\hat \theta_{\elw} - \theta_0 )
\convergeto \mathcal{N}(0, \Sigma_{\elw0})$, where
$
\Sigma_{\elw0}
=
V^{-1} \{ B_{ \dot{\ell} \dot{\ell}} - (B_{ \dot{\ell} 1} B_{ \dot{\ell} 1}^\T)/(B_{11} - \alpha_0^{-1}) \}V^{-1}
$
and
$B_{ \dot{\ell} 1} = \e \{ \dot{\ell} (Z)/ \varphi(Z ) \}$.
\een
\end{theorem}

Because $\Sigma_{\ipw} -\Sigma_{\elw} = V^{-1} B_{ \dot{\ell} h} B_{hh}^{-1}
B_{ \dot{\ell} h}^\T V^{-1}$ is a nonnegative-definite matrix,
the ELW estimator is asymptotically more efficient than the IPW estimator.
This finding remains true even if we ignore constraint \eqref{constr-h},
or if no auxiliary information is incorporated in the ELW estimator.
It can also be verified that $
\Sigma_{\elw0}- \Sigma_{\elw}
=
V^{-1}\{ B_{ \dot{\ell} h} B_{hh}^{-1} B_{ \dot{\ell} h}^\T -
(B_{ \dot{\ell} 1} B_{ \dot{\ell} 1}^\T)/(B_{11} - \alpha_0^{-1}) \} V^{-1}$
is nonnegative-definite, which means
that incorporating auxiliary information
enhances the efficiency of the proposed ELW estimator.

\subsection{Case with negligible sampling fraction}

Thus far, we have assumed that the overall sampling fraction of the big data
is nonnegligible, i.e. $\alpha_0 \in (0, 1)$.
When the volume of the big data is huge,
it is reasonable to assume that the sampling fraction may be negligible.
\begin{assumption}
\label{negligible}
Suppose there exist a positive sequence $\{b_N\}_{N=1}^{\infty}$,
a positive function $ 0<\pi_{ *}(Z)\leq 1$,
and a positive constant $ \alpha_{1*} $
such that $b_N \rightarrow \infty$,
$b_N /N \rightarrow 0$,
$ b_N \pi(Z) \rightarrow \pi_{ *}(Z)$,
and
$ b_N\alpha_{10} \rightarrow \alpha_{1*} $ as $N\rightarrow \infty$.
\end{assumption}

Under Assumption \ref{negligible},
we have $b_N\alpha_{20} = \e\{ b_N \pi(Z) \} \rightarrow \alpha_{2*} = \e\{ \pi_{ *}(Z)\}$ as $N\rightarrow \infty$.
Define $\alpha_0 = b_N\alpha_{10} + b_N\alpha_{20}$
and
\(
\varphi(Z)= b_N \alpha_{10} +b_N \pi (Z).
\)
Then, $\alpha_0 $ and $ \varphi(Z)$ converge to
$\alpha_{*} = \alpha_{1*}+\alpha_{2*}$ and
\(
\varphi_*(Z)= \alpha_{1*} + \pi_*(Z)
\), respectively.
Because $\alpha_{10}$ and the $ \pi(Z_i)$ are prespecified,
the log-likelihood \eqref{log-el}
under Assumption \ref{negligible},
up to a constant not depending on the unknown parameters $p_i$,
is equal to $ \sum_{i=1}^N D_i \log(p_i)$.
Besides the constraints $p_i\geq 0$ and
$\sum_{i=1}^N p_i=1$,
the $p_i$ in this situation should satisfy
$\sum_{i=1}^N p_i h_{e*}(Z_i) = 0$,
where
$h_{e*}(Z)= (\varphi_*(Z) - \alpha_*, h^\T(Z) )^\T$.

The maximum EL estimator of $p_i$ is
\(
\hat p_{i*}= n^{-1} \{1+\hat\lambda_*^\T h_{e*} (Z_i) \},
\)
where $\hat\lambda_*$ is the solution to
\ba
\label{constr-h*}
\frac{1}{n} \sum_{i=1}^n \frac{h_{e*} (Z_i)}{1+\hat\lambda_*^\T h_{e*} (Z_i)} = 0.
\ea
Our ELW estimator of $\theta $ is $\hat \theta_{\elw}
=\arg\min_{\theta} \sum_{i=1}^n \hat p_{i*} \ell(Z_i, \theta) $.

\begin{theorem}
\label{asym-el-cap*} 
Suppose that Assumptions \ref{assumption-data}--\ref{negligible}
hold, the distribution of $Z$ is nondegenerate,
and that
$C_{ hh*} = \e \{ h_{e*}(Z)h_{e*}^\T(Z)/ \varphi_{*}(Z ) \}$
is positive-definite.
As $N$ goes to infinity,
$ \sqrt{N /b_N}(\hat \theta_{\elw} - \theta_0 ) \convergeto \mathcal{N}(0, \Sigma_{\elw*})$
and
$\sqrt{N/b_N}(\hat \theta_{\ipw} - \theta_0) \convergeto \mathcal{N}(0, \Sigma_{\ipw*})$,
where
$
\Sigma_{\elw*}
=
V^{-1} ( C_{ \dot{\ell} \dot{\ell}*} - C_{ \dot{\ell} h*} C_{hh*}^{-1} C_{\dot{\ell} h*}^\T )V^{-1}
$
and
$\Sigma_{\ipw*} = V^{-1} C_{ \dot{\ell} \dot{\ell}*} V^{-1}$ with
$C_{ \dot{\ell} h*} = \e \{ \dot{\ell} (Z)h_{e*}^\T(Z)/ \varphi_{*}(Z ) \}$ and
$C_{ \dot{\ell} \dot{\ell}*} = \e \{ \dot{\ell}(Z) \dot{\ell}^\T(Z)/ \varphi_{*}(Z ) \}$.

\end{theorem}

Theorem \ref{asym-el-cap*} indicates that even as the sampling fraction
tends to zero, both the ELW and IPW estimators are consistent
at the rate $\sqrt{N/b_N}$, a lower rate than $\sqrt{N}$,
and our ELW estimator is still asymptotically more efficient
than the IPW estimator.
Although the asymptotic results here are slightly different from those
in Theorems \ref{asym-ipw} and \ref{asym-el},
the variances of $\hat \theta_{\elw} $ and $\hat \theta_{\ipw} $
can always be approximated by
$\Sigma_{\rm ELW}/N$ and $\Sigma_{\rm IPW}/N$, respectively.

\section{Optimal capture--recapture sampling plan}
\label{sec:samplan}

The asymptotic efficiency of subsample-based statistical inferences
depends critically on the underlying subsampling plan.
Carefully chosen sampling plans can lead to remarkable efficiency gains
over uniform sampling,
which motivates optimal subsampling for big data.

\subsection{Ideal optimal sampling plan}

MSE  is a popular evaluation criterion for the performance of a
point estimator. For a constant matrix $Q$,
Theorem \ref{asym-el} implies that $N$ times
the MSE of $ Q \hat \theta_{\elw}$ is approximated by
\bas
N\times \mse(Q \hat \theta_{\elw})
\approx N\times Q \var( \hat \theta_{\elw})Q^\T
=
\tr[
Q V^{-1} ( B_{ \dot{\ell} \dot{\ell}} - B_{ \dot{\ell} h} B_{hh}^{-1} B_{ \dot{\ell}h }^\T )V^{-1} Q^\T].
\eas
According to Theorem
\ref{asym-el-cap*}, this approximation still holds
when the sampling fraction is negligible.
The MSEs with $Q = I$ and $V$ correspond to the A- and
L-optimality criteria, respectively.
When $Q=V$,
the MSE criterion is independent of $V$, and hence has much practical convenience.
However, $Q=I$ is preferred when we are more interested in the efficiency of the ELW estimator itself.

Recall that
\bas
B_{\dot{\ell}\dot{\ell}} =
\e \left[ \frac{ \{ \dot{\ell}(Z, \theta_0) \}^{\otimes 2}}{ \varphi(Z)}\right], \,
B_{\dot{\ell}h} =
\e \left\{\frac{\dot{\ell}(Z , \theta_0) h_e^\T(Z )}{ \varphi(Z)}\right\}, \,
B_{hh} =
\e \left[\frac{ \{ h_e (Z ) \}^{\otimes 2}}{ \varphi(Z) }\right].
\eas
Because $h_e (Z ) = (\varphi (Z )-\alpha_0, h^\T(Z))^\T $,
$\e \{\dot{\ell}(Z , \theta_0)\} = 0$,
and $\e \{h(Z)\} = 0$, we have
\bas
B_{\dot{\ell}h}
=
\e \left\{\frac{\dot{\ell}(Z , \theta_0)( -\alpha_0, h^\T(Z))}{ \varphi(Z)}\right\},
\quad
B_{hh} =
\e \left[\frac{ \{ ( -\alpha_0, h^\T(Z))^\T \}^{\otimes 2}}{ \varphi(Z) }\right] - \alpha_0 e_1^{\otimes 2},
\eas
where $e_1$ is a unit vector in which the first component is 1.
Let $\pi=(\pi_1, \ldots, \pi_N)$ with $\pi_i = \pi(Z_i) $,
and $\varphi=(\varphi_1, \ldots, \varphi_N)$,
where $\varphi_i=1- (1-\alpha_{10})( 1-\pi_i)$.
Given the sampling plan $\pi$, natural
consistent ``estimators'' of
$B_{\dot{\ell}\dot{\ell}}, B_{\dot{\ell}h}$, and $ B_{hh}$
are
\bas
\hat B_{\dot{\ell}\dot{\ell}} =
\frac{1}{N} \sum_{i=1}^N \frac{ \{ \dot{\ell}(Z_i, \theta_0) \}^{\otimes 2}}{ \varphi_i}, \,
\hat B_{\dot{\ell}h} =
\frac{1}{N} \sum_{i=1}^N \frac{\dot{\ell}(Z_i, \theta_0) b_i^\T}{ \varphi_i},
\,
\hat B_{hh} =
\frac{1}{N} \sum_{i=1}^N \frac{ b_i^{\otimes 2}}{ \varphi_i}
- \alpha_0 e_1^{\otimes 2},
\eas
where $b_i= ( -\alpha_0, h^\T(Z_i))^\T$ for $i = 1,\dots, N$.
Accordingly,
a natural consistent ``estimator'' of $N\times\mse(\hat \theta_{\elw})$ is
\bas
H_*(\varphi)
&=&
\tr \{ Q V^{-1} ( \hat B_{\dot{\ell}\dot{\ell}}
-
\hat B_{\dot{\ell}h} \hat B_{hh}^{-1 }\hat B^\T_{\dot{\ell}h} ) V^{-1} Q^\T\} \\
&=&
\frac{1}{N} \sum_{i=1}^N \| a_i(\theta_0) \|^2 \varphi_i^{-1}
-
\frac{1}{N}\tr\left\{ \left( \sum_{i=1}^N \varphi_i^{-1} a_i(\theta_0) b_i^\T \right)\right.\\
&&
\left.
\times
\left( \sum_{i=1}^N ( \varphi_i^{-1} b_i b_i^\T - \alpha_0 e_1^{\otimes 2} ) \right)^{-1}
\left( \sum_{i=1}^N \varphi_i^{-1} b_i a_i^\T(\theta_0) \right) \right\},
\eas
where $a_i(\theta_0) = Q V^{-1} \dot{\ell}(Z_i, \theta_0) $.
Because there is a one-to-one map from $\pi$ to $\varphi$,
determining the optimal sampling plan $\pi$
is equivalent to determining the optimal $\varphi$.
The optimal $\varphi$ in terms of parameter estimation accuracy
is the solution to
\ba
\label{obj0}
\min_{\varphi} H_*(\varphi) \quad \mbox{s.t.}\quad
\sum_{i=1}^N \varphi_i = N\alpha_0, \;
\alpha_{10} < \varphi_i <1
\mbox{ for } i=1,\dots,N.
\ea
Unfortunately, there is no closed-form solution to problem \eqref{obj0},
which makes it impractical and motivates us
to derive a nearly optimal sampling plan.

\subsection{Nearly optimal sampling plan}
\label{sec:nearly_opt}

Nearly optimal solutions to \eqref{obj0} can be
obtained 
using several techniques.
First, we replace the objective function
$
H_*(\varphi) $
by
$ H(\varphi)$, where
\bas
H(\varphi)
=
\frac{1}{N} \sum_{i=1}^N \frac{\| a_i(\theta_0) \|^2 } { \varphi_i }
-
\frac{1}{N}\tr\left\{ \left( \sum_{i=1}^N \frac{ a_i(\theta_0) b_i^\T}{ \varphi_i } \right)
\left( \sum_{i=1}^N \frac{ b_i b_i^\T}{ \varphi_i } \right)^{-1}
\left( \sum_{i=1}^N \frac{ b_i a_i^\T(\theta_0)}{ \varphi_i } \right) \right\}.
\eas
The fact that
$
H_*(\varphi)
\leq H(\varphi)
$ holds
for any $\varphi$
implies that
a sampling plan with a small $H(\varphi)$
somehow leads to a small $H_*(\varphi)$.

Second, we transform the constrained optimization problem
\eqref{obj0} by substituting $H(\varphi)$ for $H_*(\varphi)$.
Define
\bas
H_1(\varphi, K) =
\frac{1}{N} \sum_{i=1}^N \frac{\| a_i(\theta_0) - K b_i \|^2}{ \varphi_i },
\eas
where
$K$ is a matrix of the same dimensions as $a_i(\theta_0) b_i^\T $.
Because $H_1(\varphi, K)$ is a convex function of $(\varphi, K)$,
it follows that
\bas
\min_{\varphi} H(\varphi)
=
\min_{\varphi} \min_{K } H_1(\varphi, K)
=
\min_{\varphi, K} H_1(\varphi, K),
\eas
and that
the solution to \eqref{obj0} can be approximated
by solving
\ba
\label{pi-K}
\min_{\varphi, K} H_1(\varphi, K) \quad \mbox{s.t.}\quad
\sum_{i=1}^N \varphi_i = N\alpha_0, \;
\alpha_{10} < \varphi_i<1
\mbox{ for } i=1,\dots,N.
\ea
 
If we discard the inequality constraints and retain the equality constraint, then
\bas
\min_{\varphi, K} H_1(\varphi, K) = \min_{K } \{ \min_{\varphi} H_1(\varphi, K) \}
= \min_K \frac{ \{ H_2(K) \}^2}{N^2 \alpha_0} = \frac{ \{ \min_K H_2(K) \}^2}{N^2 \alpha_0},
\eas
where
$
H_2(K ) = \sum_{i=1}^N \| a_i(\theta_0) - K b_i\|.
$
Denote $\hat K = \arg \min_K H_2(K)$.
In this situation, a nearly optimal $\varphi$ is $\hat \varphi = (
\hat \varphi_{1}, \ldots, \hat \varphi_{N})$ with
\ba
\label{optimal-pi-0}
\hat \varphi_{i}
=
\alpha_0 \cdot
\frac{ \| a_i(\theta_0) - \hat K b_{i} \|}{ N^{-1}\sum_{j=1}^N \| a_j(\theta_0) - \hat K b_{j} \|}.
\ea
Note that $(\hat \varphi, \hat K)$ is generally different from
$(\hat \varphi_*, \hat K_*)$, which is
the minimizer of problem \eqref{pi-K},
because the optimization problems with
and without the inequality constraint $\alpha_{10} < \varphi_i<1$
($i=1,\dots, N$) are not equivalent.
From a practical perspective,
we propose to take $\hat K$ as an approximation of $\hat K_*$
and adopt the optimal sampling plan with $\varphi$ solving
\ba
\label{pi-K2}
\min_{\varphi} H_1(\varphi,\hat K) \quad \mbox{s.t.}\quad
\sum_{i=1}^N \varphi_i = N\alpha_0, \;
\alpha_{10} < \varphi_i<1
\mbox{ for } i=1,\dots,N.
\ea

By the Karush--Kuhn--Tucker condition,
the solution to \eqref{pi-K2} is
\ba
\label{optimal-pi-1}
\hat \varphi_{{\rm e}i}
=
\max\left\{ \alpha_{10}, ~
\min\left( \hat \gamma \cdot
\frac{ \| a_i(\theta_0) - \hat K b_{i} \|}{ N^{-1}\sum_{j=1}^N \| a_j(\theta_0) - \hat K b_{j} \|},
~ 1 \right)
\right\},
\ea
where the subscript ``e'' denotes that $\hat \varphi_{{\rm e}i}$ is the ``exact'' solution
to \eqref{pi-K2}, and
$ \hat \gamma > 0$ is the solution to
\bas
N^{-1}\sum_{i=1}^N
\max\left\{ \alpha_{10}, ~
\min\left( \gamma \cdot
\frac{ \| a_i(\theta_0) - \hat K b_{i} \|}{ N^{-1}\sum_{j=1}^N \| a_j(\theta_0) - \hat K b_{j} \|},
~ 1 \right)
\right\} = \alpha_0.
\eas

\subsection{Practical considerations}
\label{sec:prac}

The sampling plans with $\hat \varphi_{i}$
and $\hat \varphi_{{\rm e}i}$ are not practically applicable, because
both of them depend on $\theta_0$, which needs to be estimated
beforehand. To this end, the convention is to draw an initial
sample, say $\{ \tilde z_i = (\tilde y_i, \tilde x_i^\T)^\T: i=1, \ldots, m \}$,
by uniformly sampling from the big data being studied.
The first capture in our capture--recapture sampling
plays exactly the same role.
Let $\tilde \theta_m = \arg\min_{\theta} \sum_{i=1}^m \ell (\tilde z_i, \theta)$
and
$\tilde V$ be a consistent estimator of $V$
based on the first-capture sample.
Denote
$
\tilde K = \arg\min \sum_{i=1}^m \| \tilde a_i - K \tilde b_i\|,
$
where
$
\tilde b_i = ( -\alpha_0, h^\T(\tilde z_i))^\T,
$ and
$
\tilde a_i =\tilde V^{-1} \dot{\ell}(\tilde z_i, \tilde \theta_m)
$ in the A-criterion
or $
\tilde a_i = \dot{\ell}(\tilde z_i, \tilde \theta_m)
$ in the L-criterion.
Calculating $\tilde K$ may be computationally intensive, and
so we use the least-squares estimate
$
\tilde K = \left( \sum_{k=1}^m \tilde a_k \tilde b_k^\T\right)
\left( \sum_{j=1}^m \tilde b_j \tilde b_j^\T\right)^{-1}
$
instead.
Define
\ba
\label{optimal-pi-1-prac}
\tilde \varphi_{{\rm e}i}
=
\max\left\{ \alpha_{10}, ~
\min\left( \tilde \gamma \cdot
\frac{ \| \tilde a_i - \tilde K \tilde b_{i} \|}{ m^{-1}\sum_{j=1}^m \| \tilde
a_j - \tilde K \tilde b_{j} \|}, ~ 1 \right)
\right\},
\ea
where $ \tilde \gamma > 0$ is the smallest solution to
\bas
m^{-1}\sum_{i=1}^m
\max\left\{ \alpha_{10}, ~
\min\left( \gamma \cdot
\frac{ \| \tilde a_i - \tilde K \tilde b_i \|}{ m^{-1}
\sum_{j=1}^m \| \tilde a_j - \tilde K \tilde b_{j} \|},
~ 1 \right)
\right\}
= \alpha_0.
\eas
Our recommended sampling plan
for the second capture is $\tilde \pi=(\tilde \pi_1,
\ldots, \tilde \pi_N)$ with $ \tilde \pi_i =
( \tilde \varphi_{{\rm e}i} - \alpha_{10})/(1-\alpha_{10})$,
where $\alpha_{10} \in (0, 1)$
is the known sampling fraction of the first capture.

For the models in Table \ref{tab:example} (except the quantile regression model),
the matrix $V$ can be consistently estimated by the moment estimation method
based on the first-capture sample.
The estimation of $V$ in the quantile regression model
is more challenging because it depends on the unknown conditional density function $f(y\mid x)$.
Following \cite{powell1989estimation}, we estimate this $V$ using the kernel estimator
$
\tilde V = (m h_m)^{-1} \sum_{i=1}^{m} K\{ (\tilde y_i - \tilde x_i^\T\tilde \theta_m)/h_m \} \tilde x_i \tilde x_i^\T,
$
where $K(\cdot)$ is a kernel function, usually chosen to be a density function,
and $h_m$ is the bandwidth.

\section{Sample size determination}\label{sec:sampsize}

For a given subsample, the performance of the IPW and ELW estimators depends not only on the underlying sampling plan,
but also on the size of the subsample.
If the size $n$ or the ideal size $n_0$ of a Poisson subsample
is too small, the resulting estimator will
be so unstable that it does not make any sense.
When the (optimal) sampling plan is fixed,
it is necessary to specify the subsample size
that guarantees the resulting estimate meets
a certain precision requirement.
To the best of our knowledge, this issue has never
been discussed in the literature of subsampling for big data.
We address the issue of determining $n_0$ under two precision requirements on $\hat \theta_{\elw}$:
 (R1)
The MSE of $\hat\theta_{\rm ELW}$ is no greater than
a prespecified positive constant $C_0$, i.e.,
$ \mse(\hat\theta_{\rm ELW})\leq C_0$.
 (R2)
The absolute error of $\hat\theta_{\rm ELW}$
is no greater than a critical value $d>0$ at the confidence level ($1 - a$), i.e.,
\ba
\label{requirement-ae}
P(\|\hat\theta_\elw - \theta_0\| \leq d) \geq 1 - a.
\ea

We assume that the sample fraction $\alpha_{10}>0$ of the first capture
is known, but
that for the second capture $\alpha_{20}$
is unknown.
Because $n_0/N = \alpha_0 = 1-(1-\alpha_{10})(1-\alpha_{20})$,
when the (optimal) sampling plan is fixed,
determining $\alpha_{20}$ is equivalent to determining $n_0$.
Recall that
a nearly optimal subsampling plan
can be approximated by \eqref{optimal-pi-0} or $\tilde\varphi_* =
(\tilde\varphi_{*1}, \dots, \tilde\varphi_{*N})$, where
$
\tilde \varphi_{*i}
$
is
$\hat \varphi_i$
with $\hat K$ replaced by $\tilde K$.
With the sampling plan $\tilde\varphi_*$,
an upper bound for the MSE of $\hat\theta_{\rm ELW}$ is
\bas
H(\tilde\varphi_*)/N
&=&
\frac{1}{N^3 \alpha_0} \left\{ \sum_{j=1}^N \| a_j(\theta_0) - \tilde K b_{j}\| \right\}^2
=
\frac{1}{n_0} \left\{ \frac{1}{N} \sum_{j=1}^N \| a_j(\theta_0) - \tilde K b_{j}\| \right\}^2,
\eas
which can be estimated by
$n_0^{-1} \{ m^{-1} \sum_{j=1}^m
\| \tilde a_j - \tilde K \tilde b_{j}\| \}^2$.
Under requirement (R1),
a sufficient approximation is to constrain
$n_0^{-1} \{ m^{-1} \sum_{j=1}^m
\| \tilde a_j - \tilde K \tilde b_{j}\| \}^2\leq C_0$.
Note that the elements of $\tilde K$ and $\widetilde b_j$
contain the unknown parameter $\alpha_0 = n_0/N$.
Therefore, the minimal sample size $n_0$ that satisfies requirement (R1)
should be the solution to
\ba
\label{eq:n_0}
n_0 = \frac{1}{C_0} \left\{ \frac{1}{m}
\sum_{j=1}^m \| \tilde a_j - \tilde K \tilde b_{j}\| \right\}^2.
\ea
This is our first recommended sample size determination method,
which we denote as M1 for convenience.

To determine the sample size under requirement (R2),
note that the inequality
$\|\hat\theta_\elw - \theta_0\| \leq d$ is equivalent to
$\zeta^\T \Sigma _\elw \zeta \leq N d^2$, where
$\zeta = \sqrt{N}\Sigma _\elw ^{-1/2}(\hat\theta_\elw - \theta_0)$
approximately
follows the $q$-dimensional standard normal distribution, where $q$
is the dimension of $\theta$.
The distribution of $\zeta^\T \Sigma _\elw \zeta$ can be
further approximated by a weighted chi-square distribution of
$\sum_{k=1}^q \lambda_k \zeta_k^2$, where
the $\lambda_k$ are the eigenvalues of $\Sigma _\elw$ and
the $\zeta_k$ are i.i.d. standard normal random variables.
According to \cite{kim2006analytic}[Lemma 2, page 453],
the cumulative distribution of $\sum_{k=1}^p\lambda_k \zeta_k^2$ can be approximated by
that of $\nu^{-1} \chi^2_{\nu}$,
where $\nu = \sum_{k=1}^q \lambda_k/\sum_{j=1}^q \lambda_j^2 $.
It follows that
$ P(\|\hat\theta_\elw - \theta_0\| \leq d)
\approx P(\chi^2_{\nu} \leq \nu N d^2 ),
$
which together with \eqref{requirement-ae} implies the approximation
$
\nu N d^2 = \chi^2_{ \nu}(1 -a),
$
where $\chi^2_{ \nu}(1 - a)$ is the
$(1 - a)$th quantile of the chi-square
distribution with $ \nu$ degrees of freedom.

Moreover, $\nu$ is approximately equal to
$\tilde\nu = \sum_{k=1}^q \tilde\lambda_k/\sum_{j=1}^q \tilde\lambda_j^2 $,
where
the $\tilde\lambda_k$ are the eigenvalues of
$\tilde\Sigma _\elw =
\tilde V^{-1} ( \tilde B_{ \dot{\ell} \dot{\ell}} - \tilde B_{ \dot{\ell} h}\tilde B_{hh}^{-1}
\tilde B_{ \dot{\ell} h}^\T )\tilde V^{-1}$.
Herein, $\tilde B_{ \dot{\ell} \dot{\ell}}$, $\tilde B_{ \dot{\ell} h}$,
and $\tilde B_{hh}$ are the sample-mean estimates of
$B_{ \dot{\ell} \dot{\ell}}$, $B_{ \dot{\ell} h}$,
and $B_{hh}$ based on the first-capture sample.
Because
$\Sigma _\elw$ (and hence $\lambda_k$) depends on $\alpha_0 = n_0/N$,
so do $\tilde\Sigma _\elw$, $\tilde\lambda_k$, and $\tilde\nu$.
We denote $\tilde\nu$ by $\tilde \nu(n_0)$ to highlight this dependence.
Our recommended sample size $ n_0$ under requirement (R2), denoted as M2,
is the root of
\ba
\label{eq:n_0_accu}
\tilde \nu(n_0) = \nu_*,
\ea
where $\nu_*$ is the solution to $ \nu N d^2 = \chi^2_{ \nu}(1 -a)$ with respect to $\nu$.

\section{Simulations}\label{sec:simu}

In this section, we present the results of simulations to evaluate the
finite-sample performance of the proposed estimation and sampling strategy
and the sample size determination method.

\subsection{Simulation settings}
We generate a big dataset of size $N=50,000$
from each of the following three examples,
corresponding to Poisson regression, binomial regression, and quantile regression models.

\begin{example}[Poisson regression] 
Given $X$, $Y$ follows a Poisson regression model
with $\e(Y|X) = \exp(X^\T \theta_0)$ and
$\theta_0 = -0.5\times (1, 1, 1,1,1,1,1)^\T$.
Four scenarios are considered to generate the
covariates $X_i=(X_{i1}, \ldots, X_{i7})^\T$:
Case 1. $X_{ij}$ are i.i.d. from $U(0, 1)$,
the standard uniform distribution;
Case 2.
$X_{ij} $ for $j\neq 2$ and $\varepsilon_i$ are i.i.d. from $U(0, 1)$,
and take $X_{i2} = X_{i1} + \varepsilon_i$.
In this case, the correlation coefficient of
$X_{i1} $ and $X_{i2} $ is around 0.7.
Case 3.
The same setting as case 2, except that
$ \varepsilon_i \sim U(0, 0.1)$.
In this case, the correlation coefficient of
$X_{i1} $ and $X_{i2} $ is around 0.995.
Case 4.
The same setting as case 2, except that
$ X_{i6}$ and $ X_{i7}$ are i.i.d. from $U(-1, 1)$.
In this case, the covariates have different supports.

\end{example}

\begin{example}[Logistic regression]
The settings here are the same as those in
Example 1, except that
$Y$ given $X$ follows a logistic regression model
with mean
$\e(Y|X) = \exp(\theta_0^\T X)/\{1 + \exp(\theta_0^\T X)\}$.
\end{example}

\begin{example}[Quantile regression] 
Given $X$, $Y$ follows a linear regression model $Y = \beta_0^\T X + \epsilon$,
where $X = (1, X_2, \dots, X_5)^\T$,
$X_2, \dots, X_5$ are i.i.d. from N(0,1),
$\beta_0 = -0.5\times (1, 1, 1,1,1)^\T$,
and the error distribution is to be specified.
Given $\tau \in (0, 1)$,
$Q_\tau(Y\mid X) = \beta_0 ^\T X + Q_\tau(\epsilon)$,
where $Q_\tau(\epsilon)$ is the $\tau$th quantile of $\epsilon$,
and
$\theta_0 = (Q_\tau(\epsilon)-0.5,-0.5,-0.5,-0.5,-0.5)^\T$.
We consider four combinations of error distribution and $\tau$:
Case 1. $ \mathcal{N}(0,1)$ and $\tau = 0.5$;
Case 2. $ \mathcal{N}(0,1)$ and $\tau = 0.75$;
Case 3. $ |\mathcal{N}(0,1)|$ and $\tau = 0.5$;
Case 4. $ |\mathcal{N}(0,1)|$ and $\tau = 0.75$.

\end{example}

We take the response mean of the big data as auxiliary information.
Let ELW and ELWAI denote the ELW methods without and with
the auxiliary information, together with
the corresponding nearly optimal capture--recapture sampling plan.
We compare the performance of ELW and ELWAI with UNIF,
the usual M-estimation with one-step uniform sampling,
and 
the IPW method together with
the corresponding optimal sampling plan.
In Examples 1 and 2,
the MV subsampling probabilities
of \cite{yu2020optimal} are used in IPW,
while in Example 3,
IPW is chosen to be
the OSQR  of \cite{fan2021optimal},
which is also an IPW-based method.
Note that
a shrinkage technique
was used with a tuning parameter $\varrho$
when calculating the MV optimal subsampling probabilities
in \cite{yu2020optimal}[equation (21)].
For consistency with the setup of \cite{yu2020optimal},
we fix $\varrho = $ 0.2 in our numerical studies.
If an initial sample (the first capture) is required for a method,
we fix its average sample size $r_0$ to be 200.
We consider the average size $r$ of the second sample (the recapture)
to be 300, 500, 700, 1000, 1200, 1500, 1700, and 2000, respectively.
To ensure a fair comparison, we set the average sample size to $r_0 + r$ for
UNIF.

\subsection{Comparison of estimation efficiency}\label{eq:sim-eff}

\begin{figure}[h]
\centering
\includegraphics[width= 3.7cm]{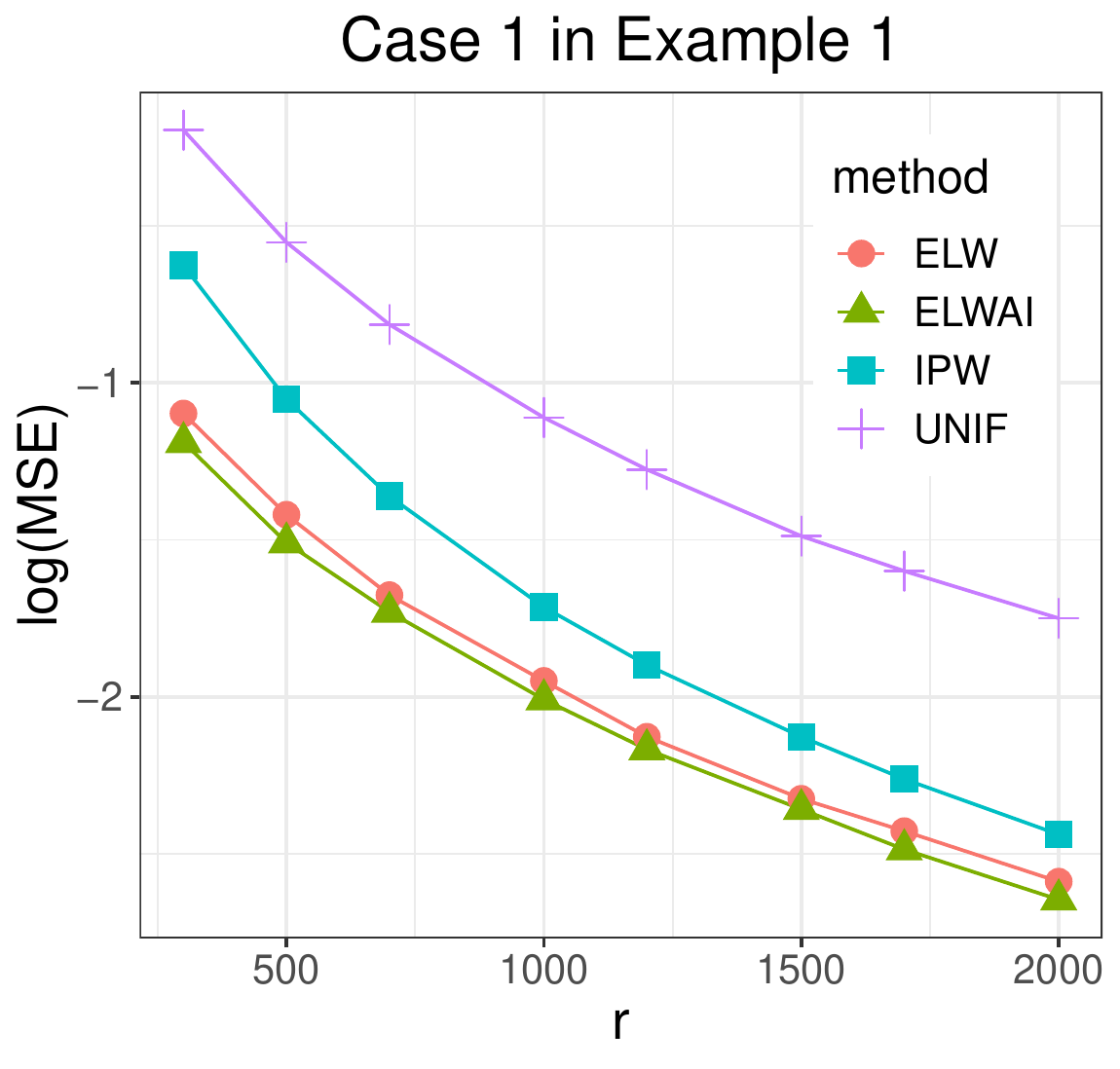}
\includegraphics[width= 3.7cm]{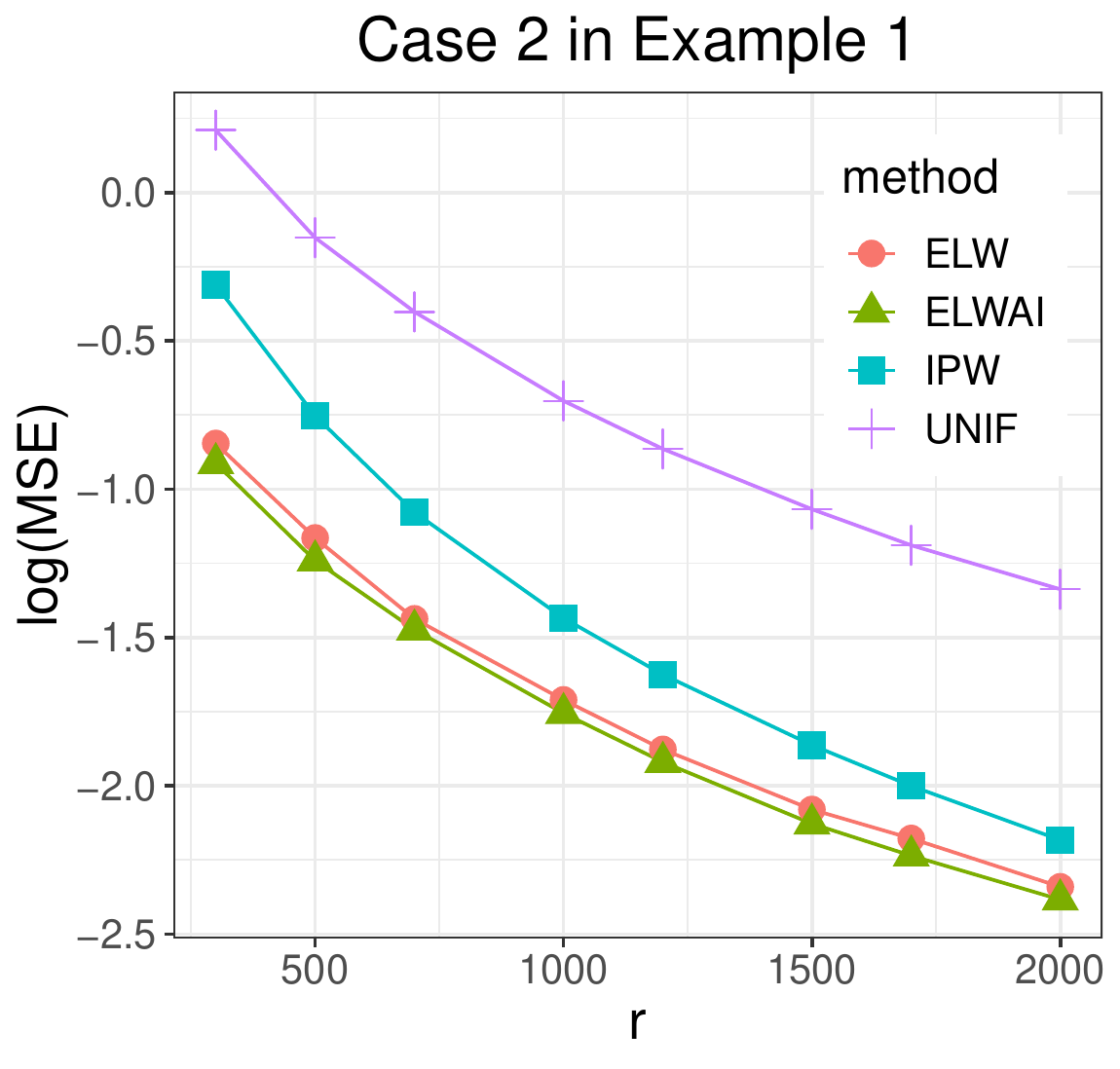}
\includegraphics[width= 3.7cm]{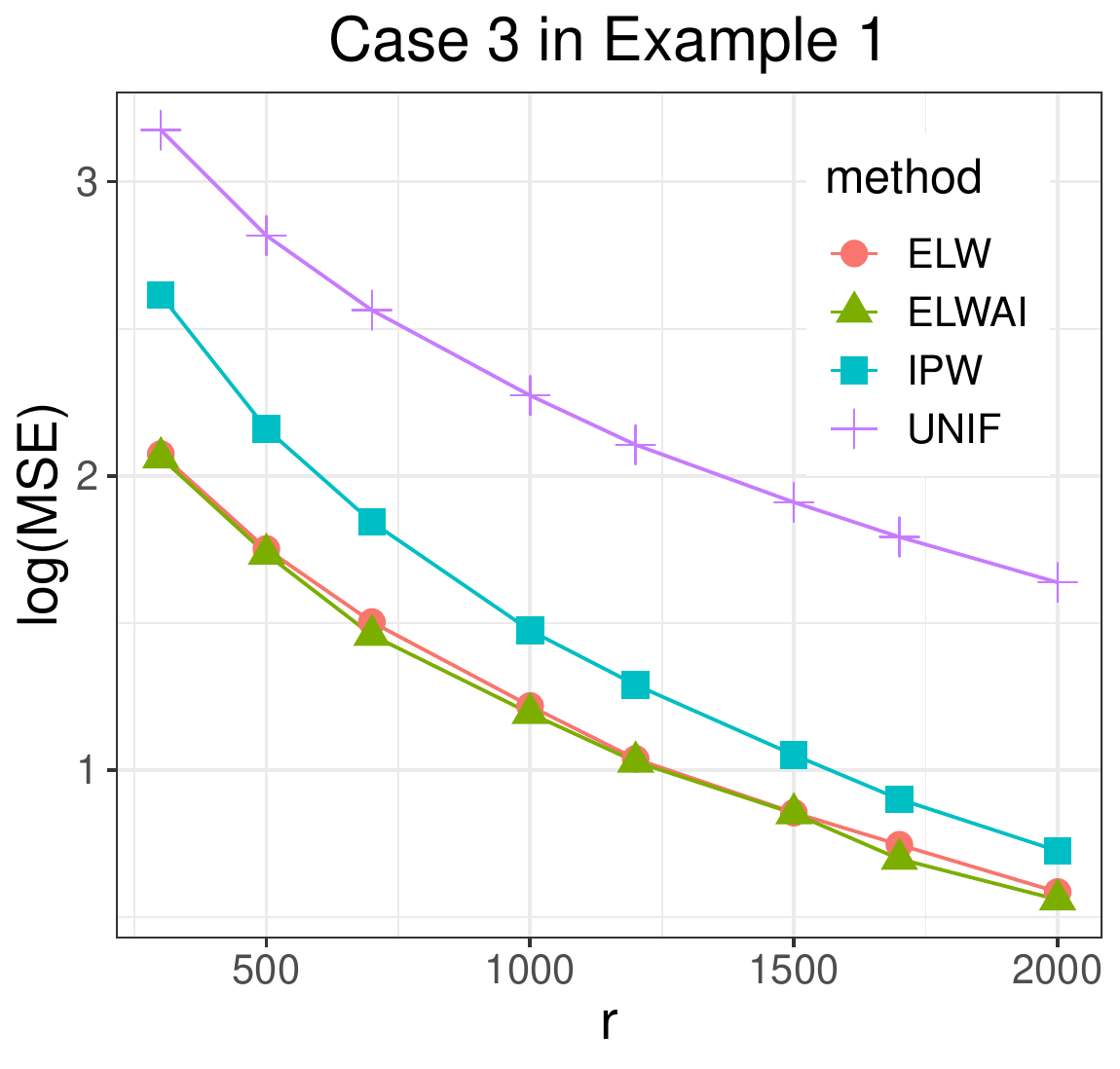}
\includegraphics[width= 3.7cm]{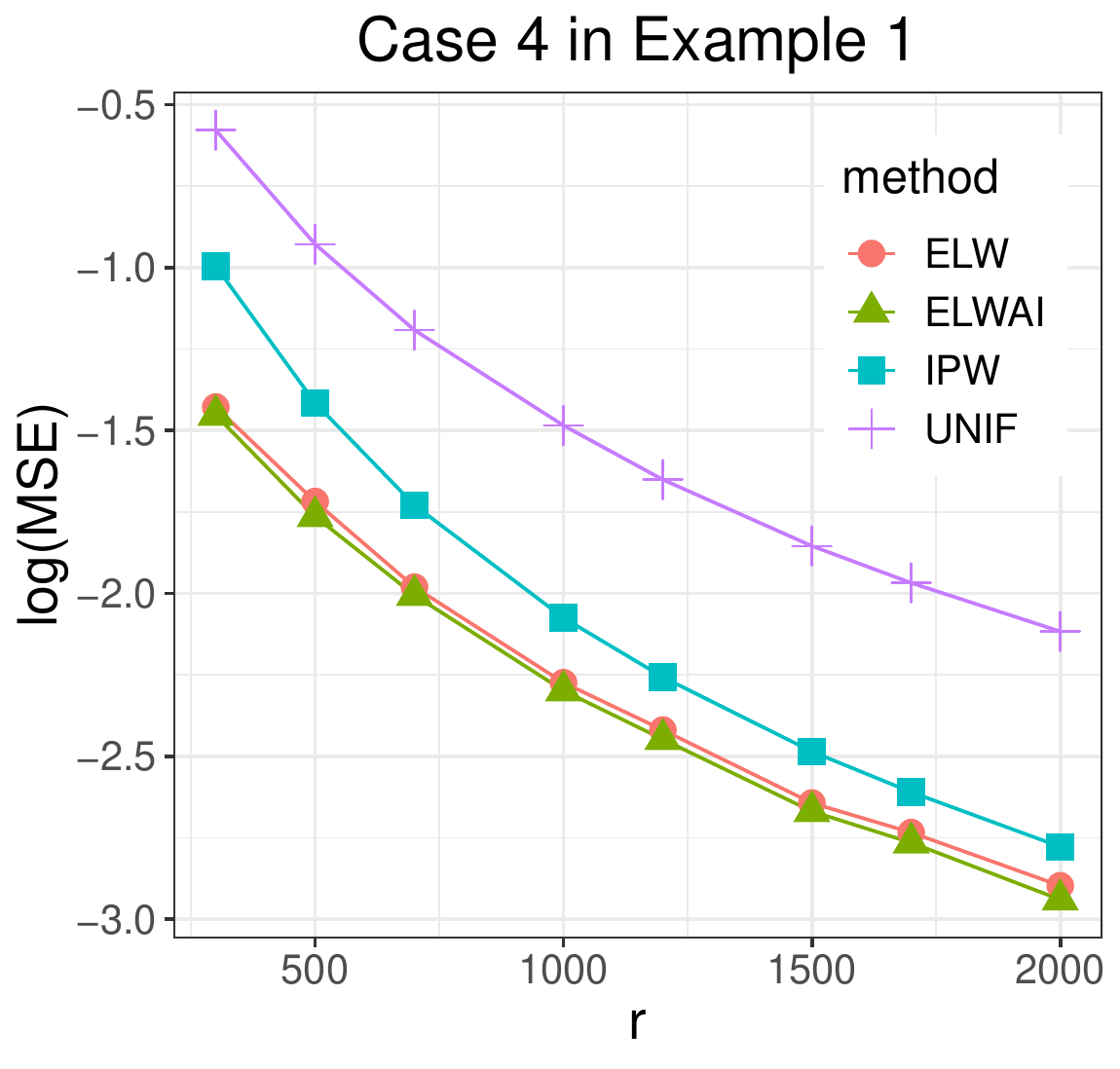}\\
\includegraphics[width= 3.7cm]{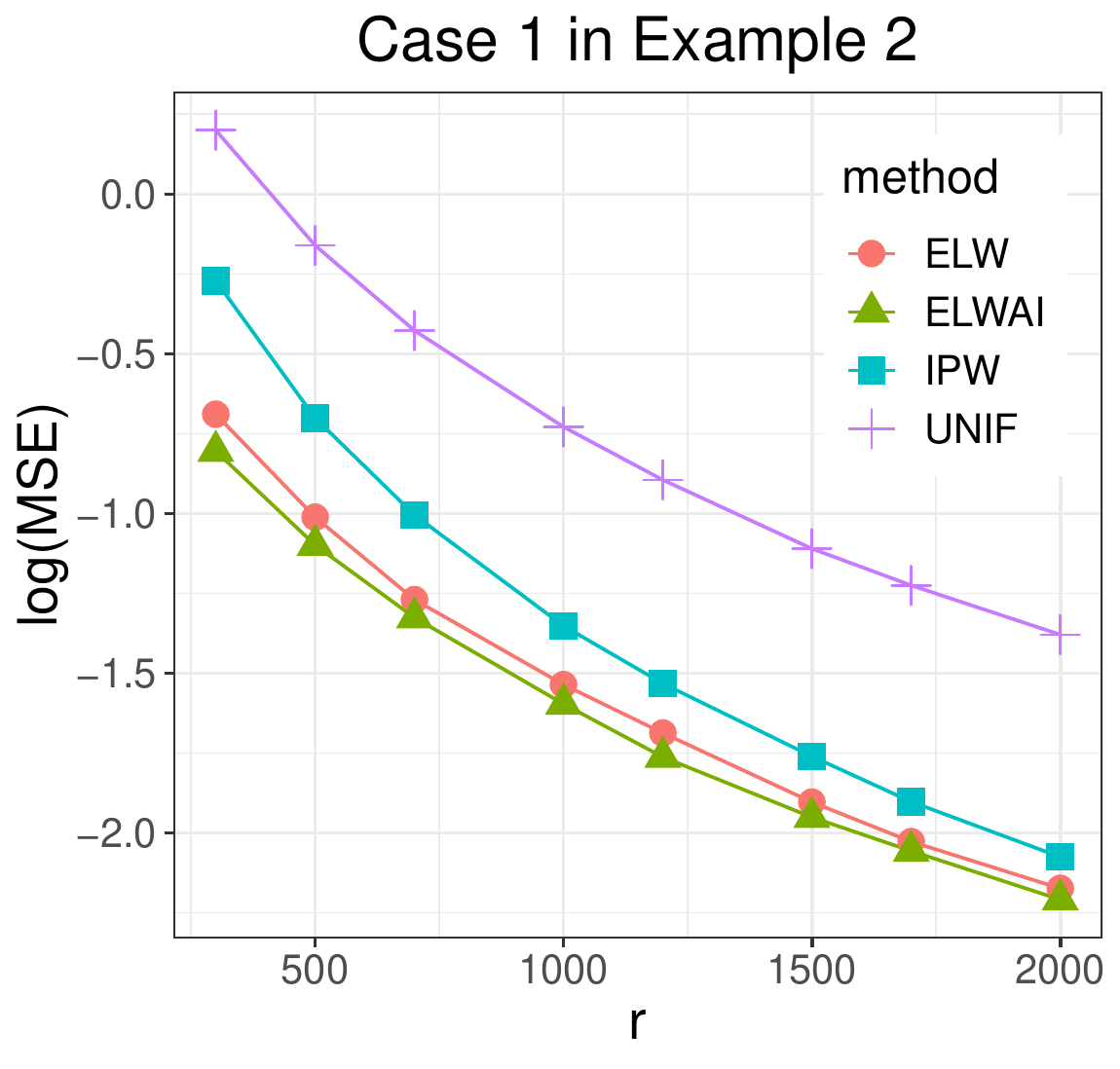}
\includegraphics[width= 3.7cm]{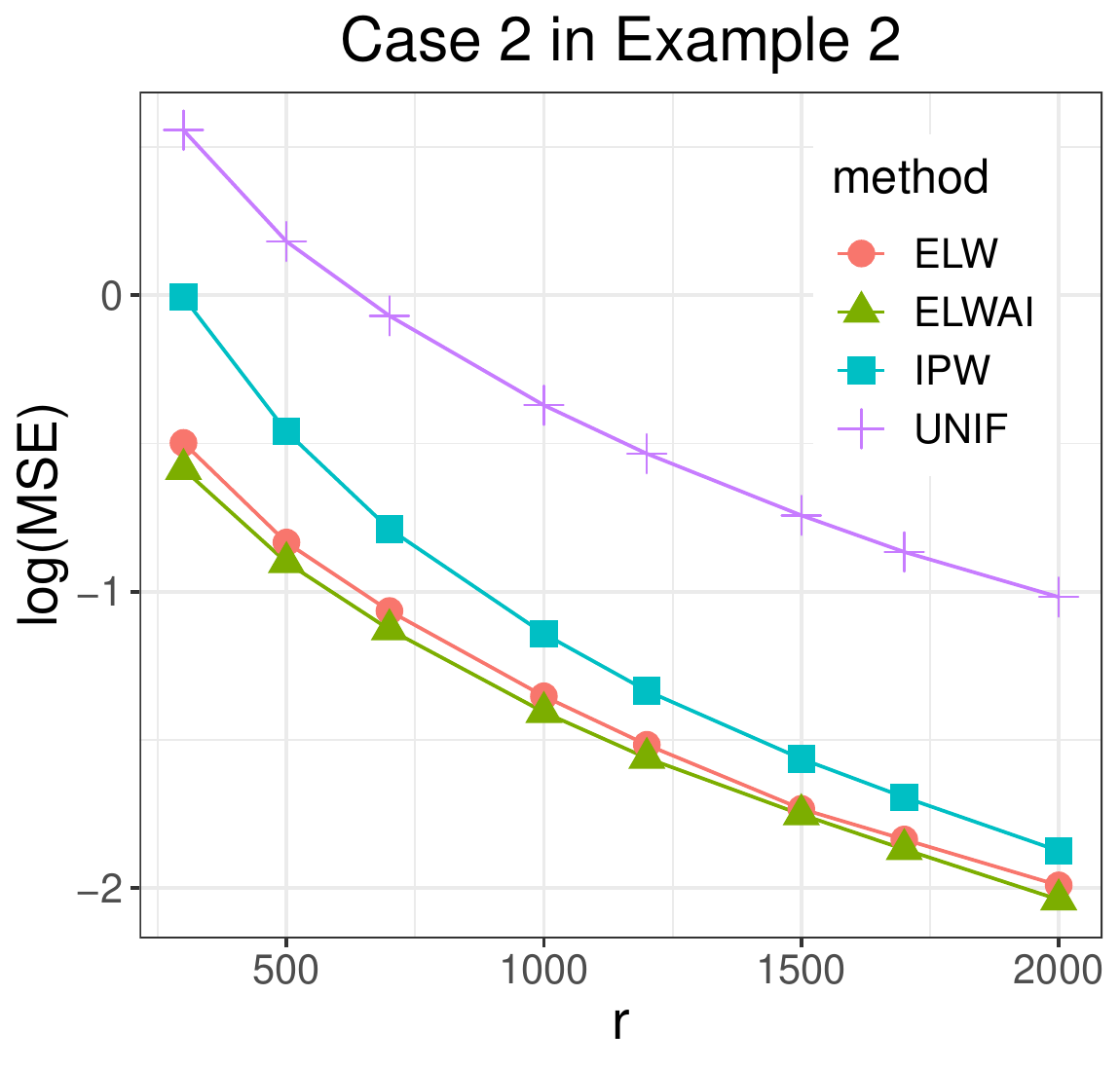}
\includegraphics[width= 3.7cm]{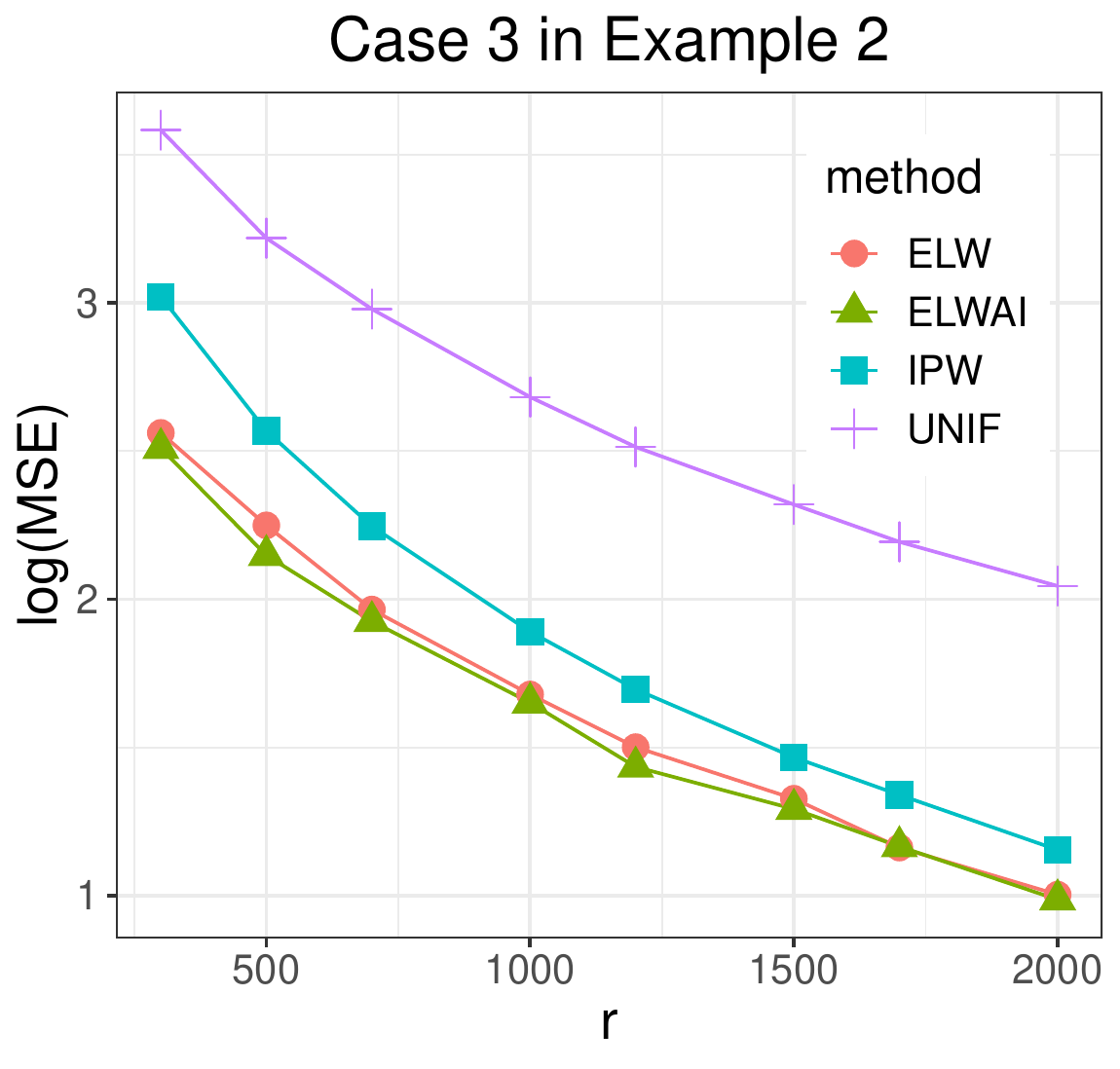}
\includegraphics[width= 3.7cm]{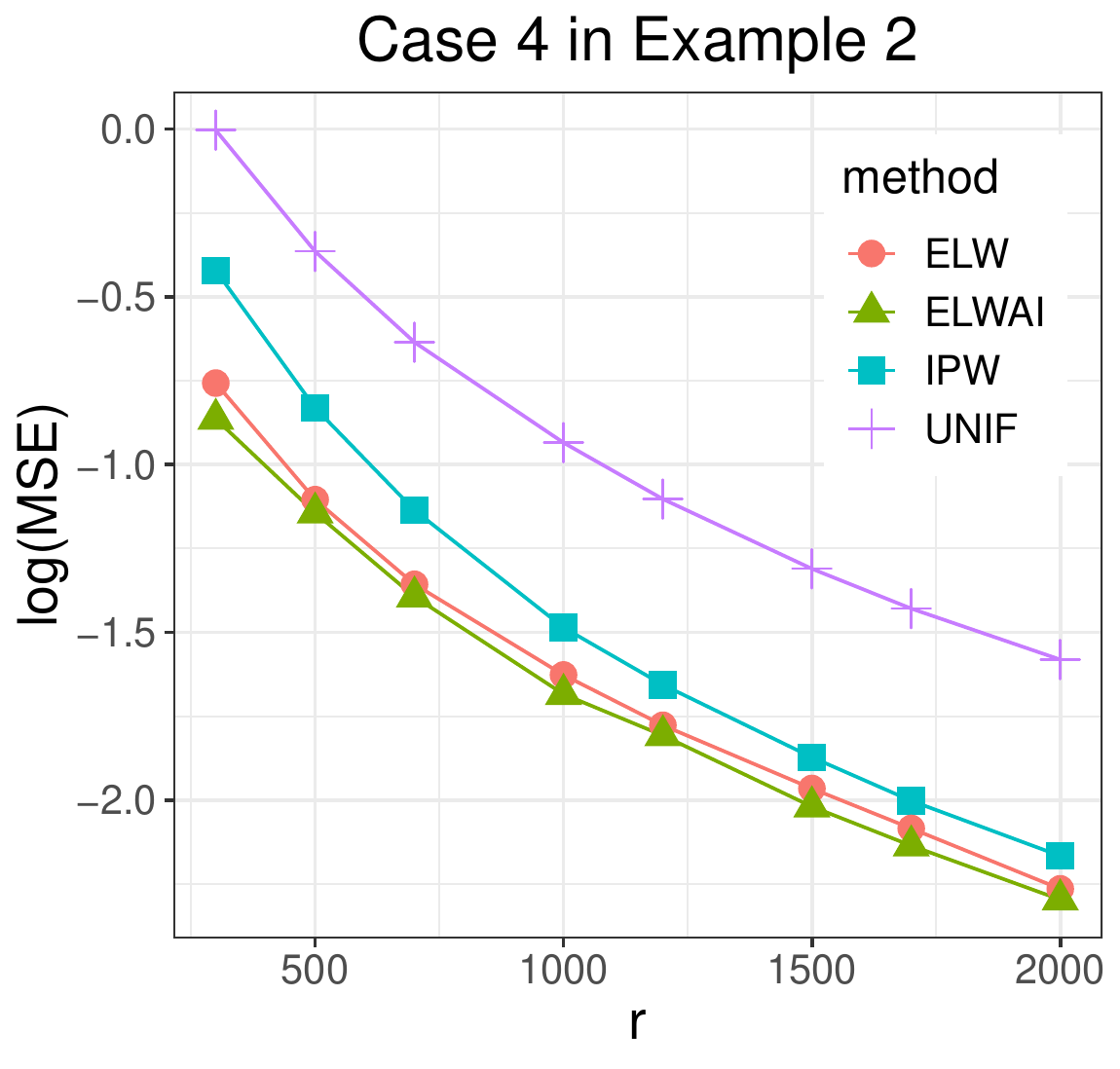}\\
\includegraphics[width= 3.7cm]{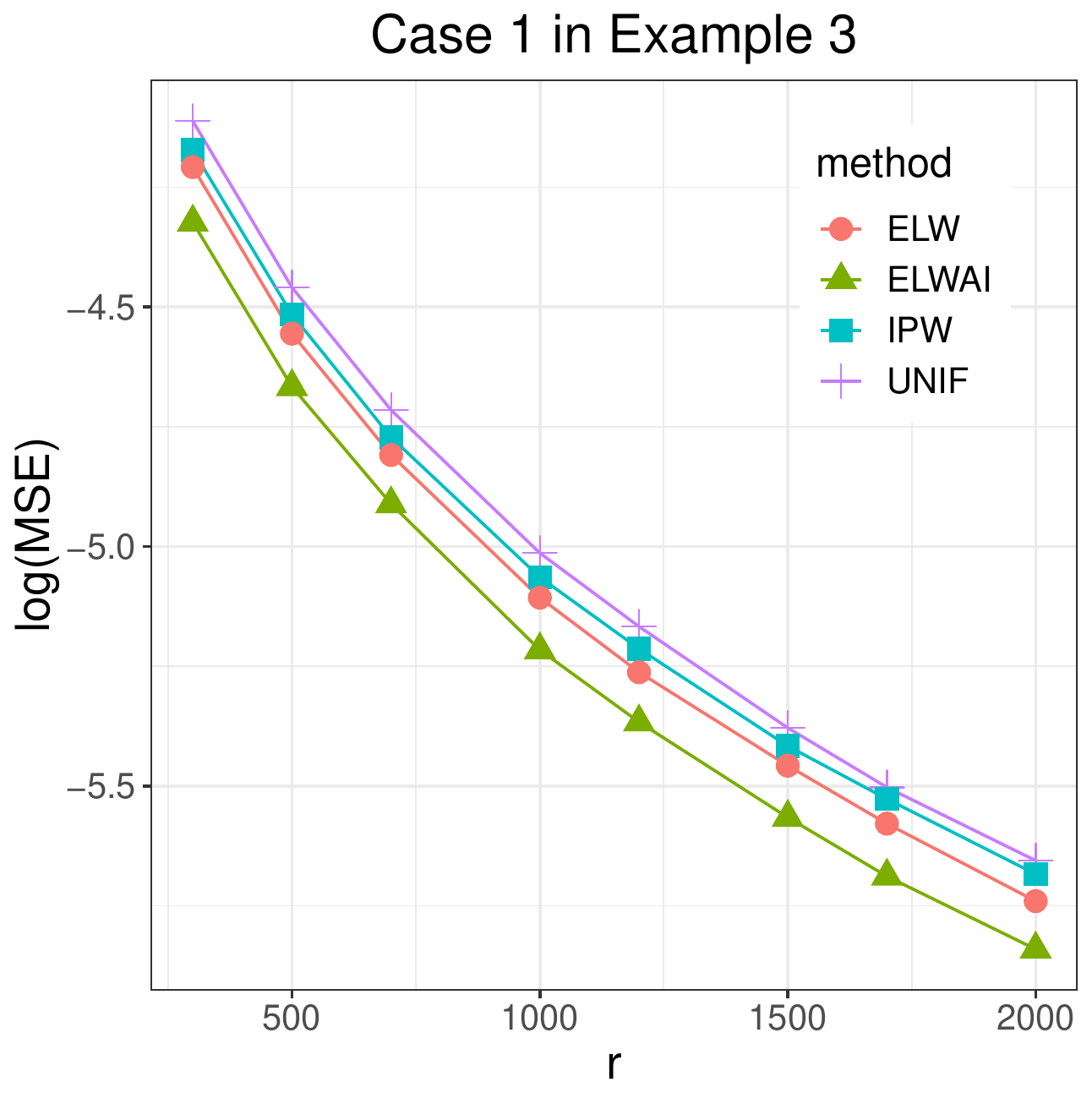}
\includegraphics[width= 3.7cm]{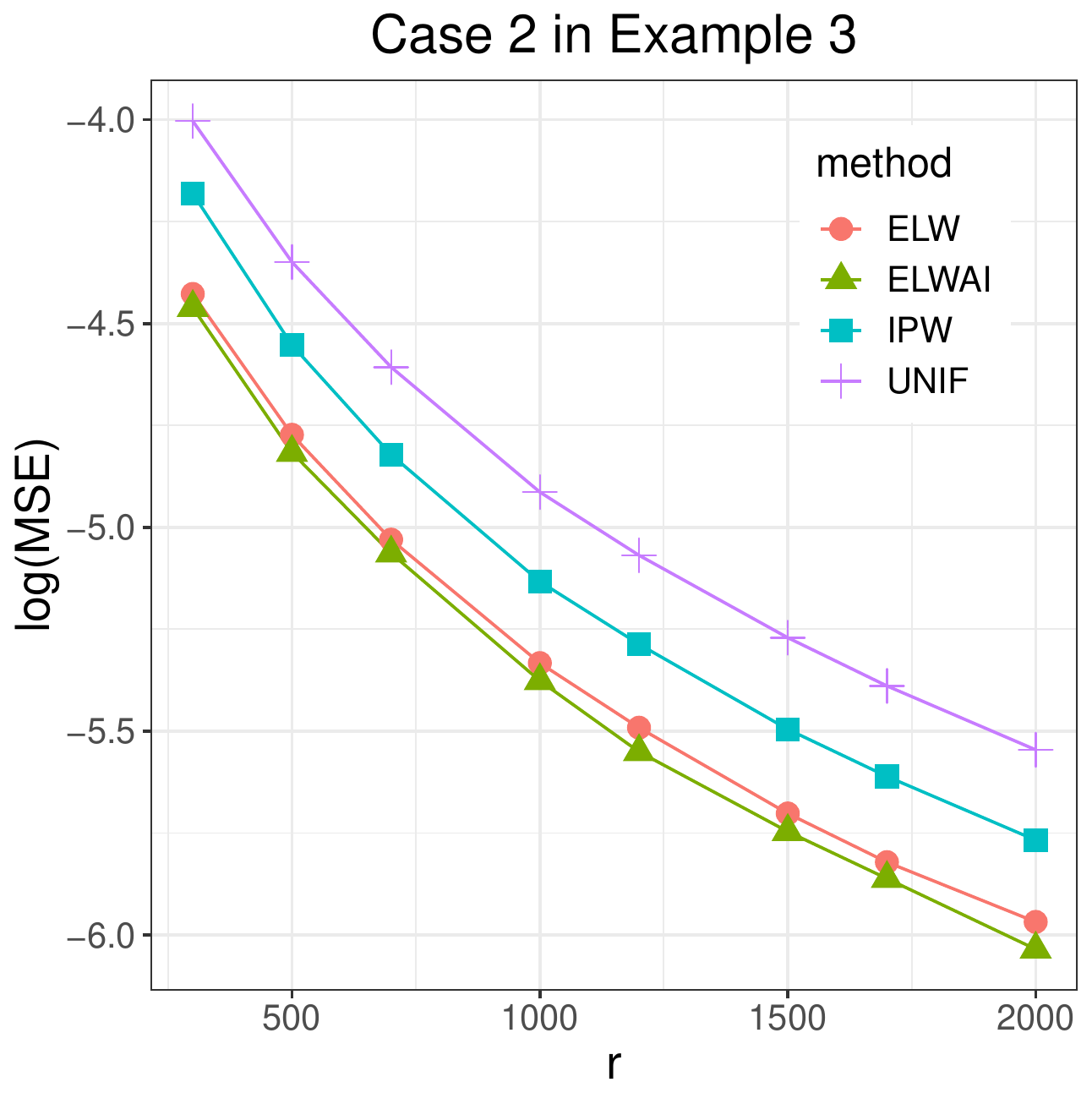}
\includegraphics[width= 3.7cm]{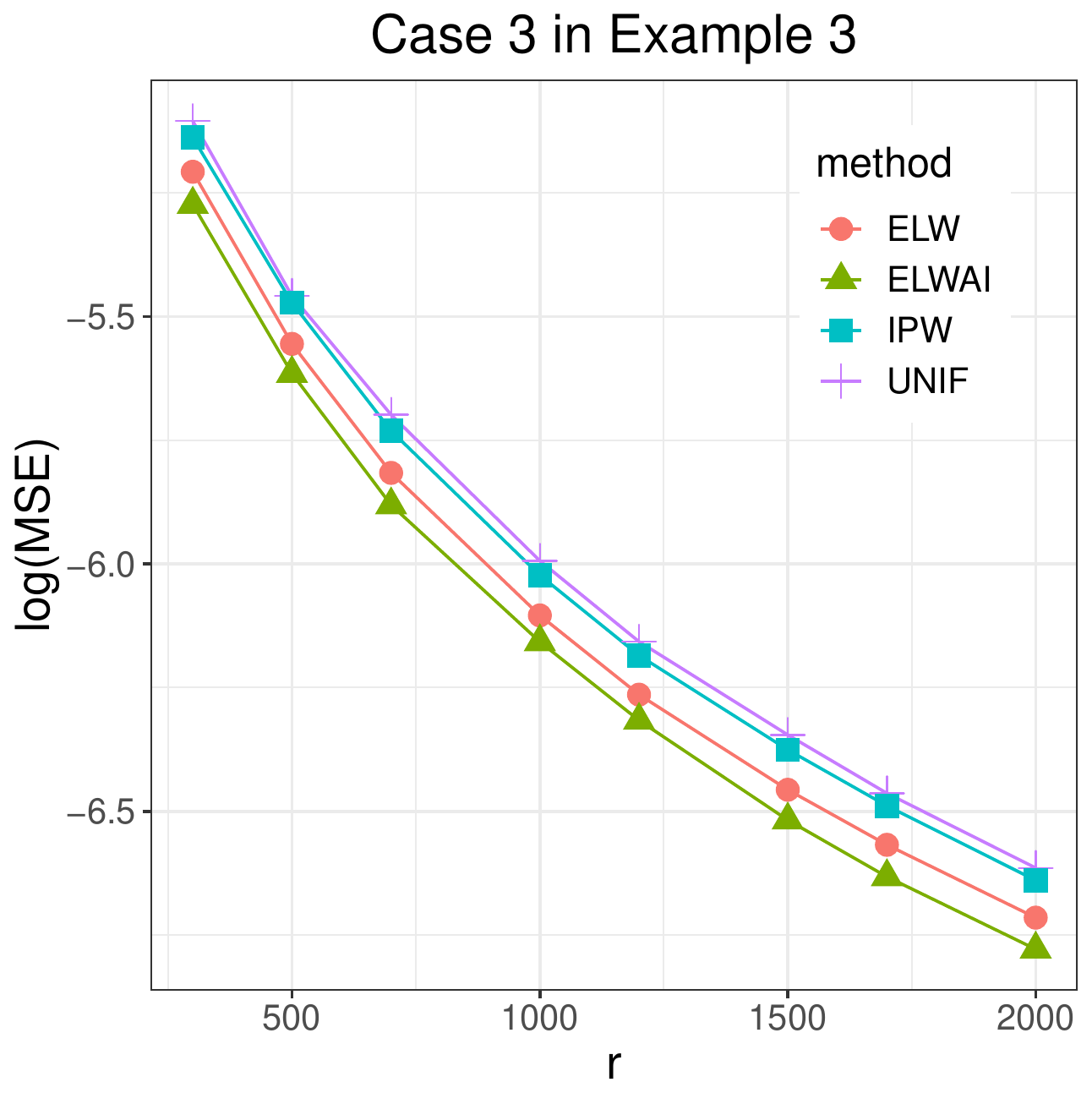}
\includegraphics[width= 3.7cm]{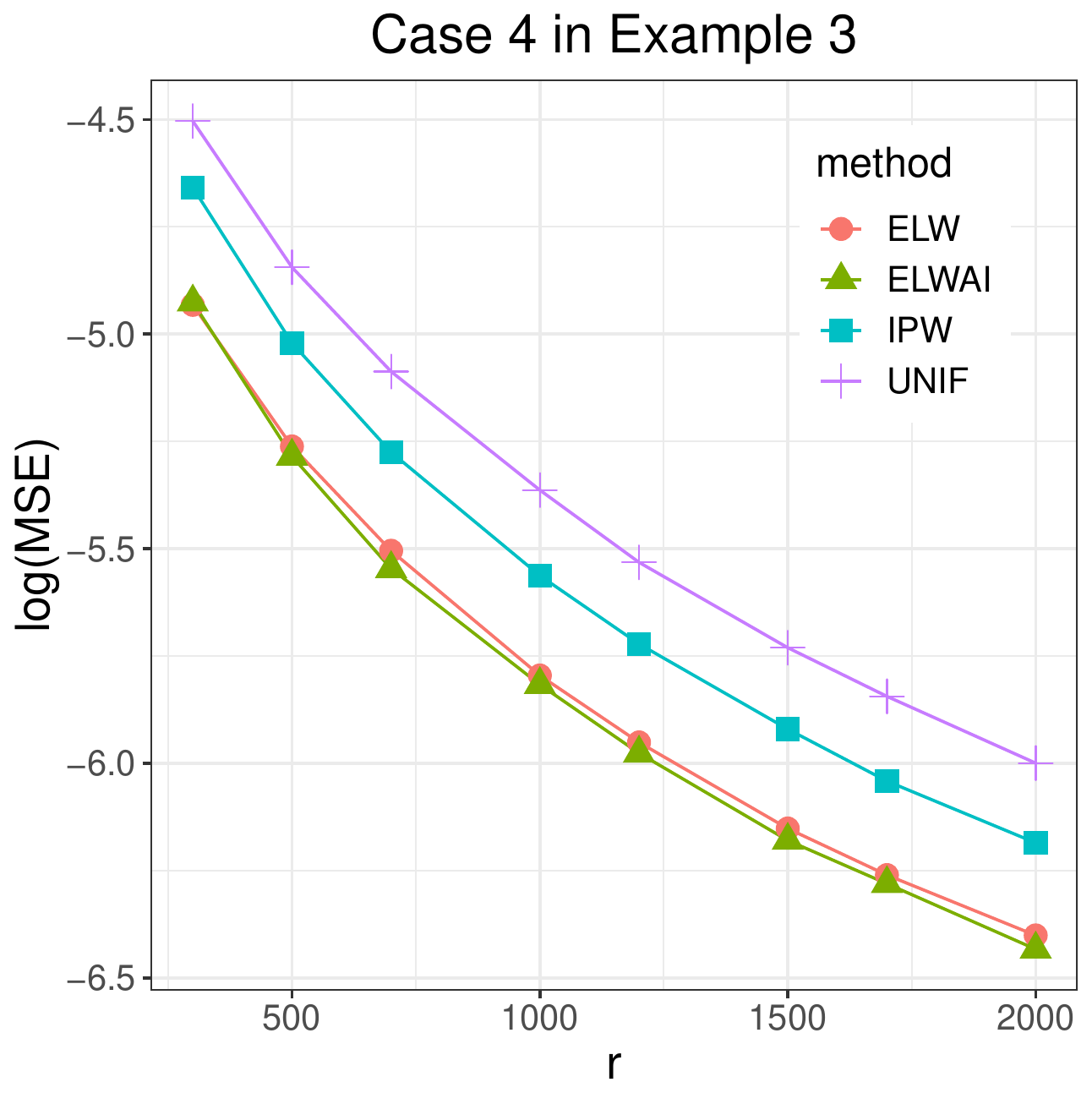}
\caption{ Plots of the logarithm of MSE versus $r$ for
UNIF, IPW, ELW, and ELWAI under the A-criterion.}
\label{fig:sim-MSE-A}
\end{figure}

\begin{figure}[h]
\centering
\includegraphics[width= 3.7cm]{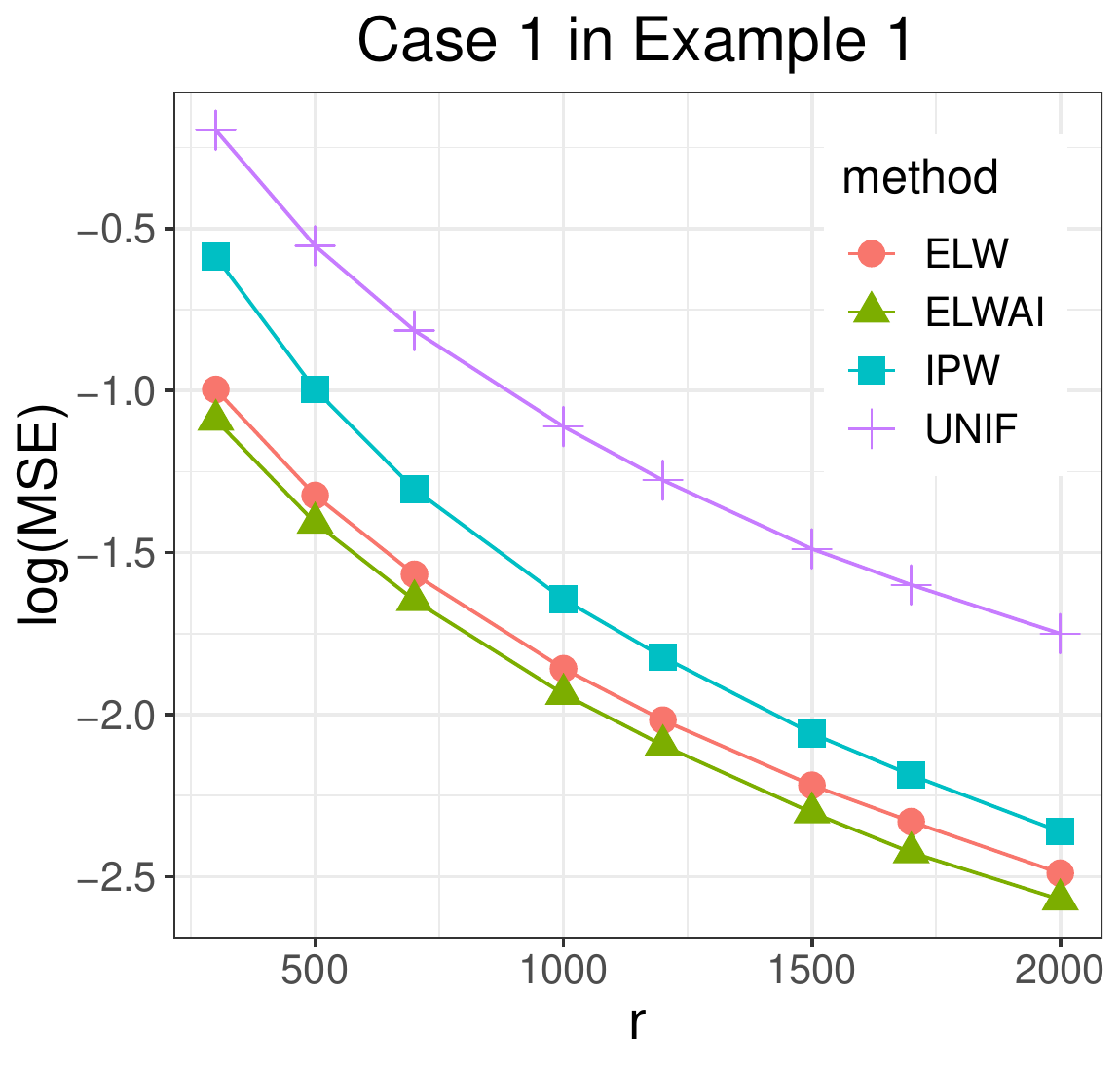}
\includegraphics[width= 3.7cm]{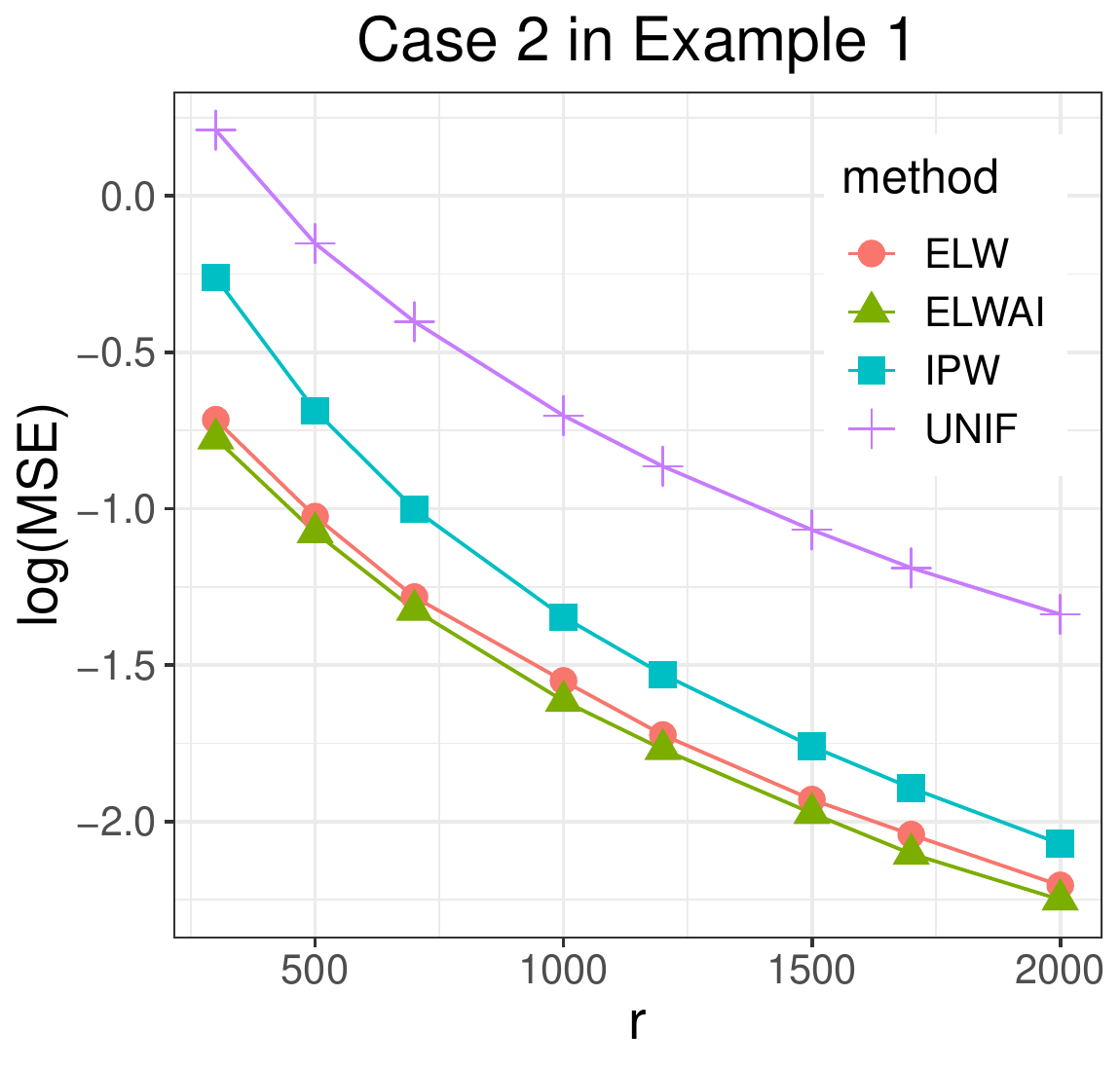}
\includegraphics[width= 3.7cm]{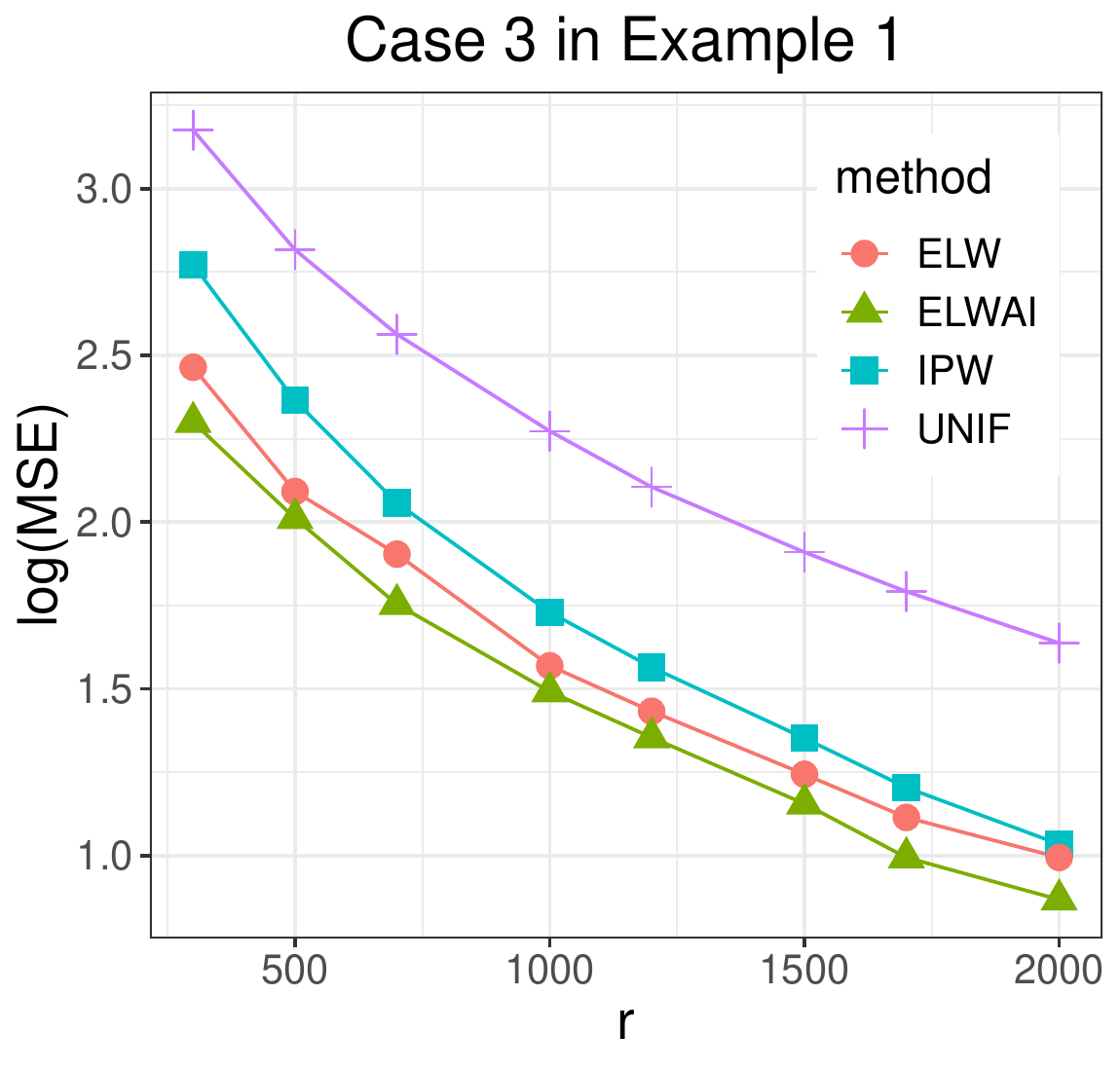}
\includegraphics[width= 3.7cm]{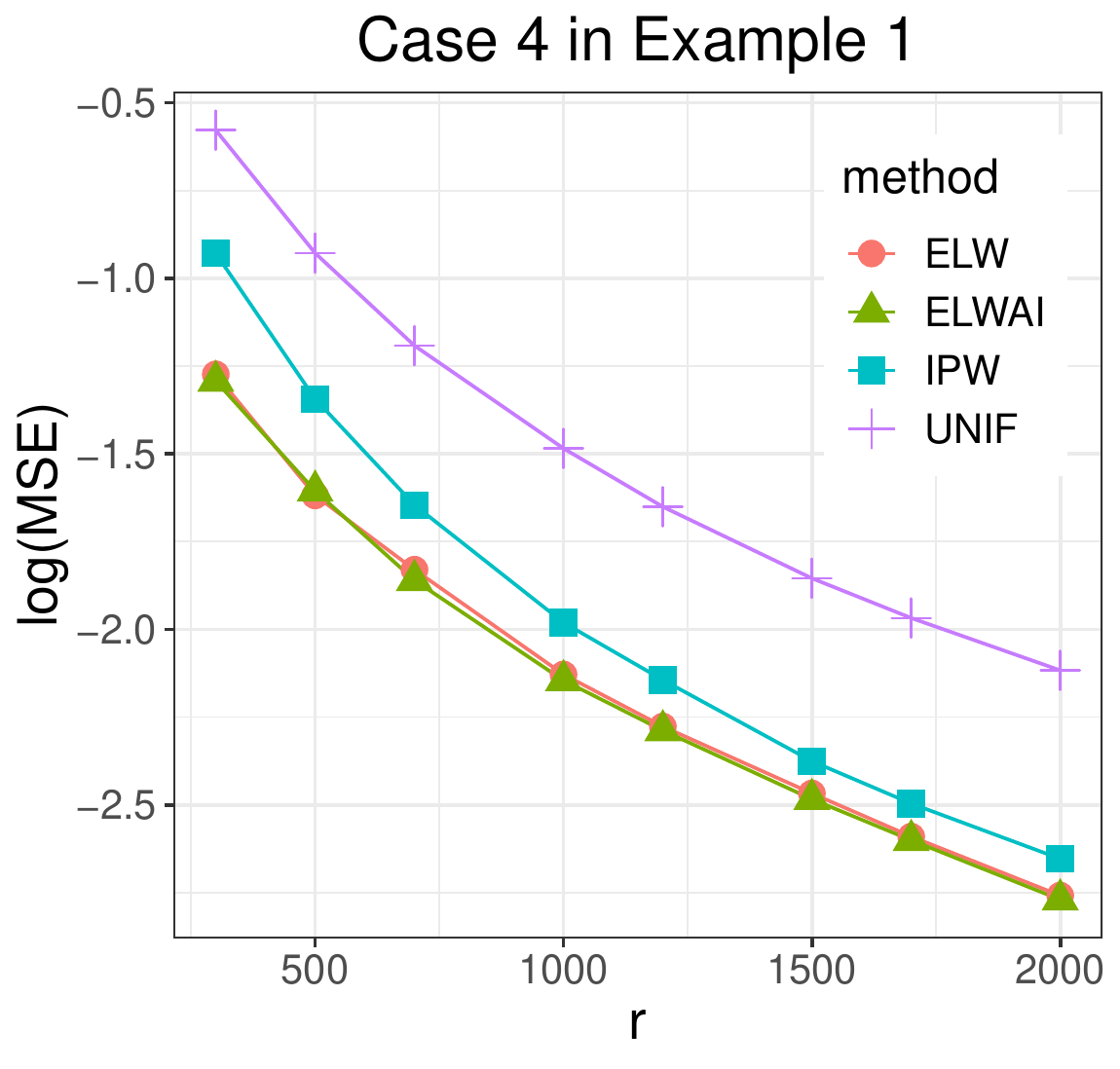}\\
\includegraphics[width= 3.7cm]{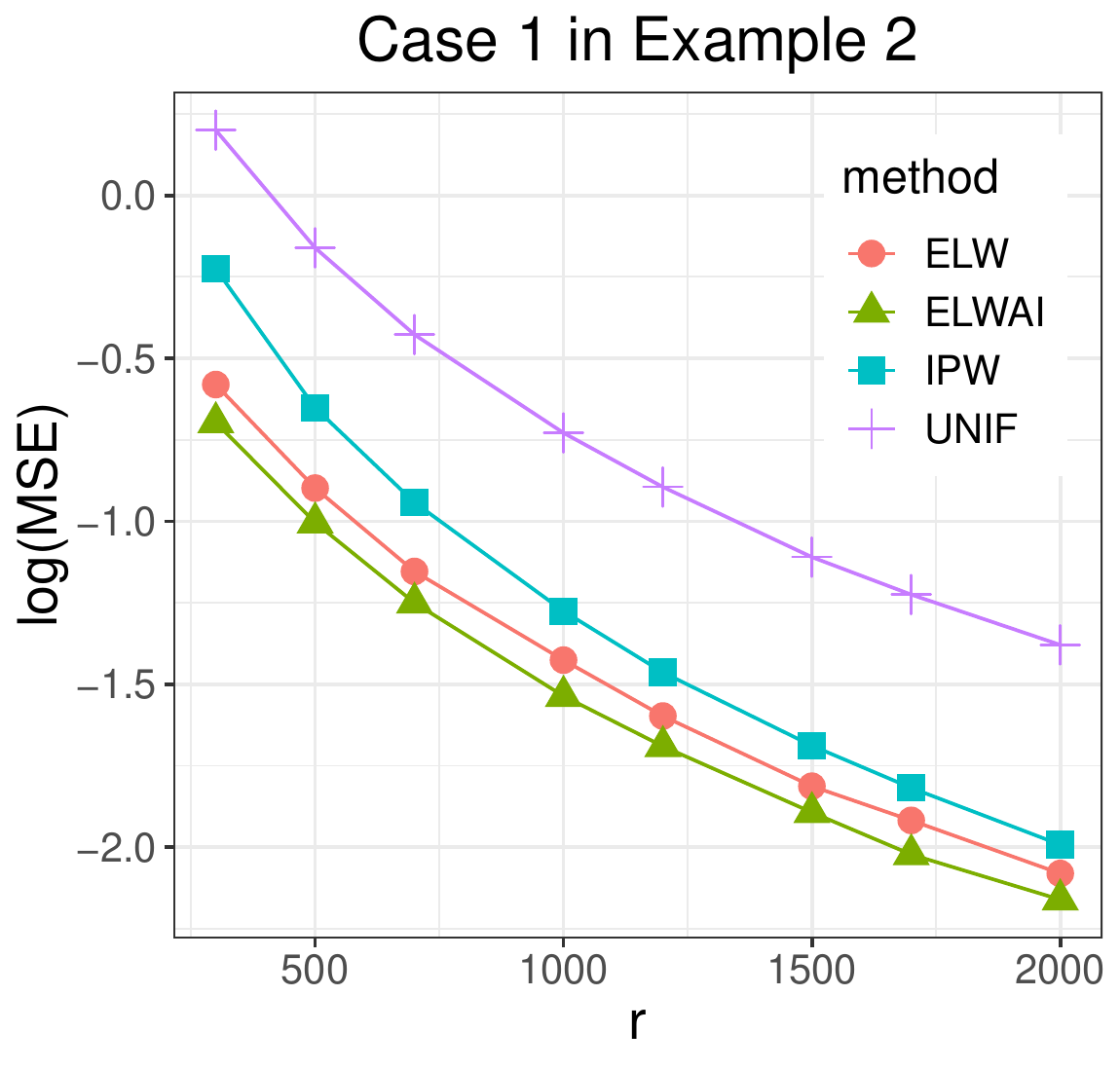}
\includegraphics[width= 3.7cm]{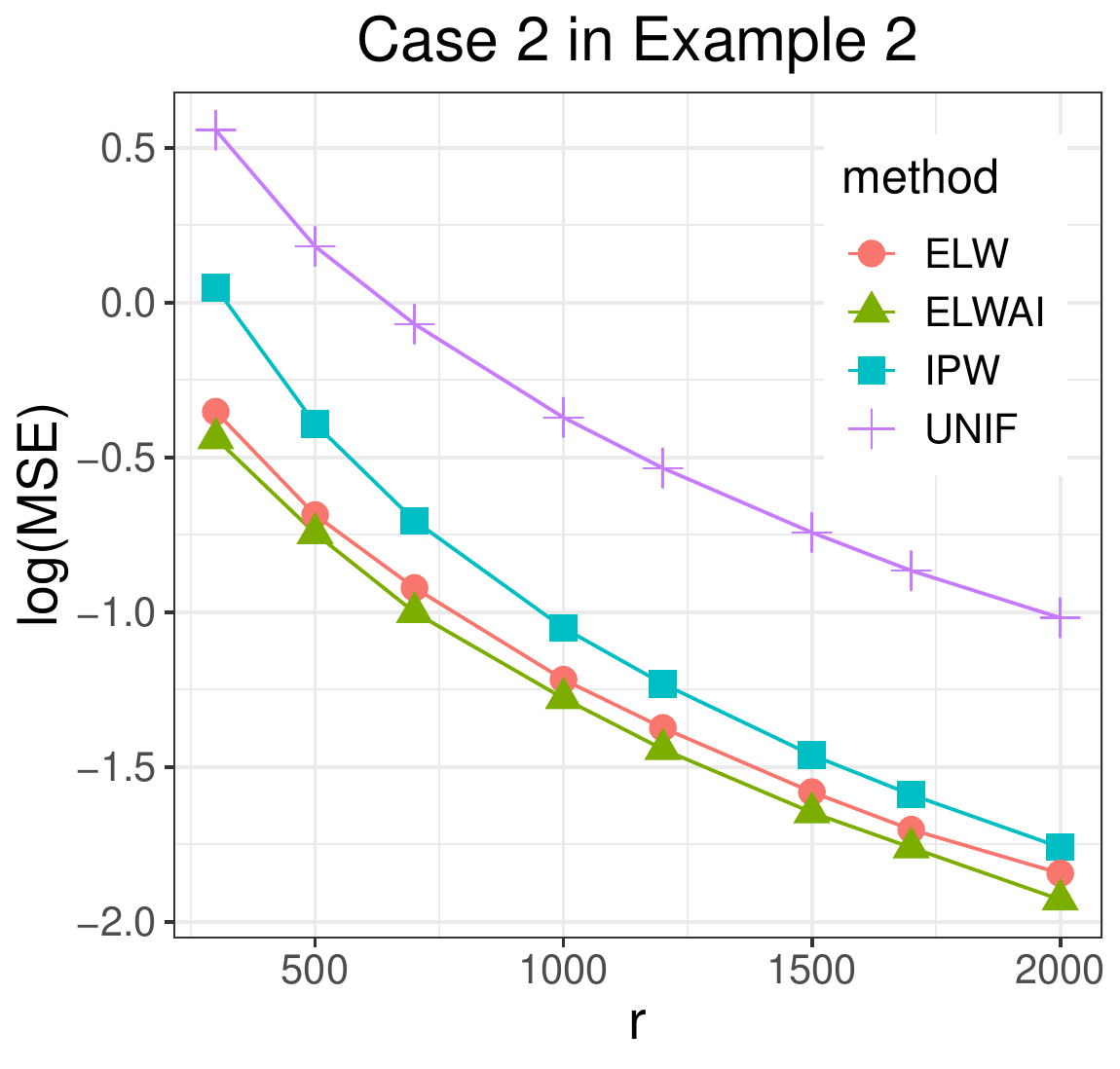}
\includegraphics[width= 3.7cm]{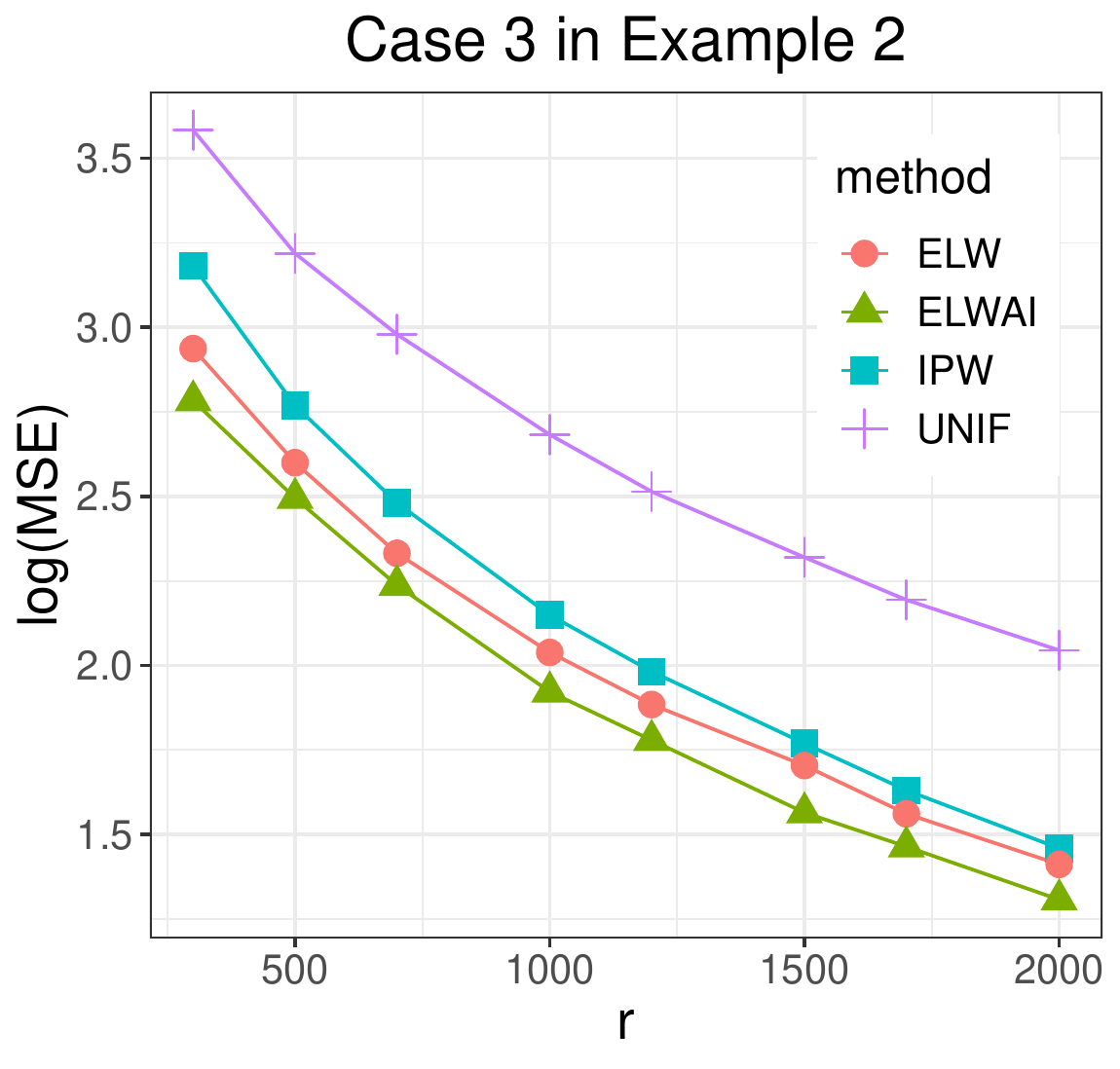}
\includegraphics[width= 3.7cm]{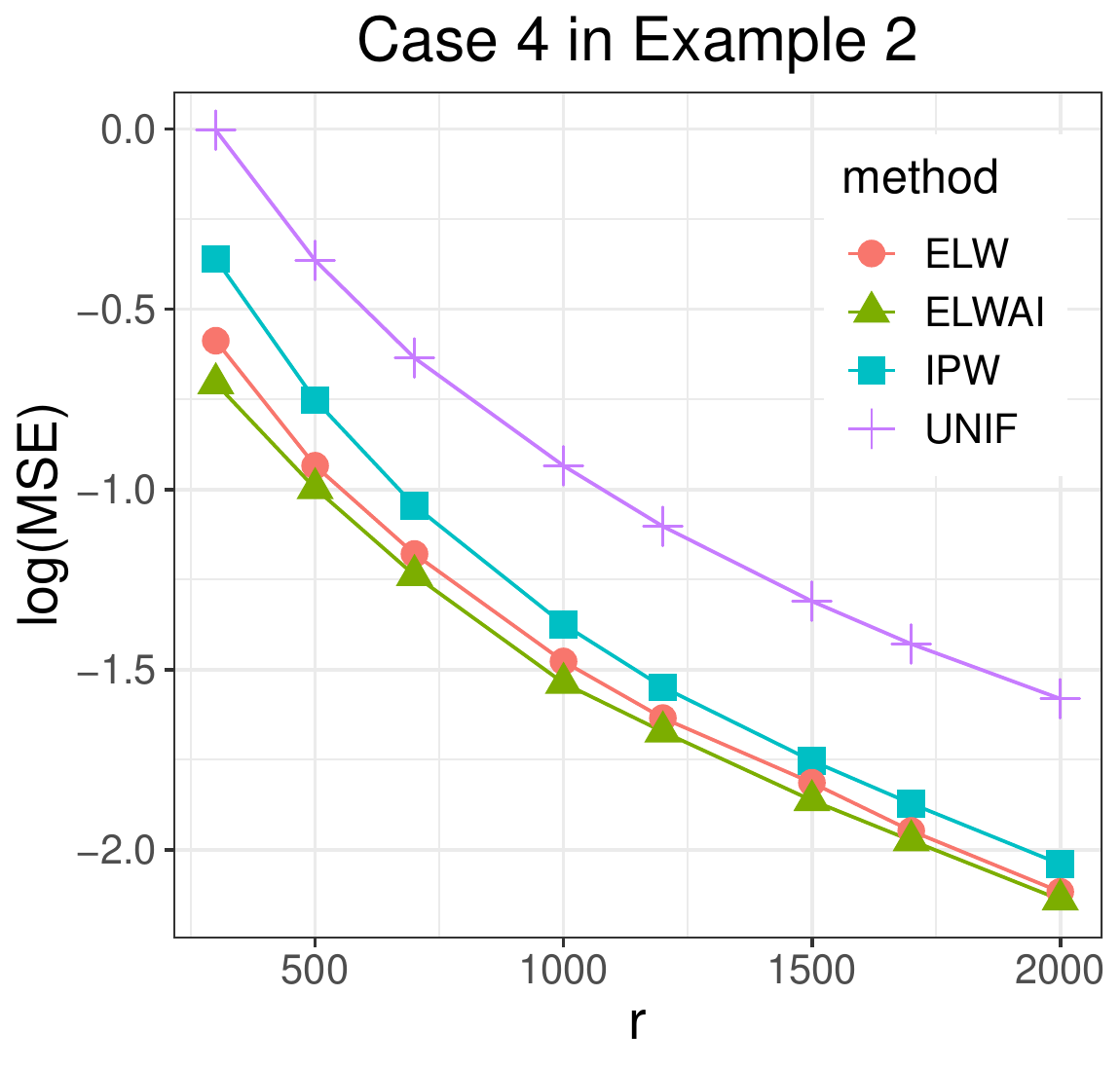}\\
\includegraphics[width= 3.7cm]{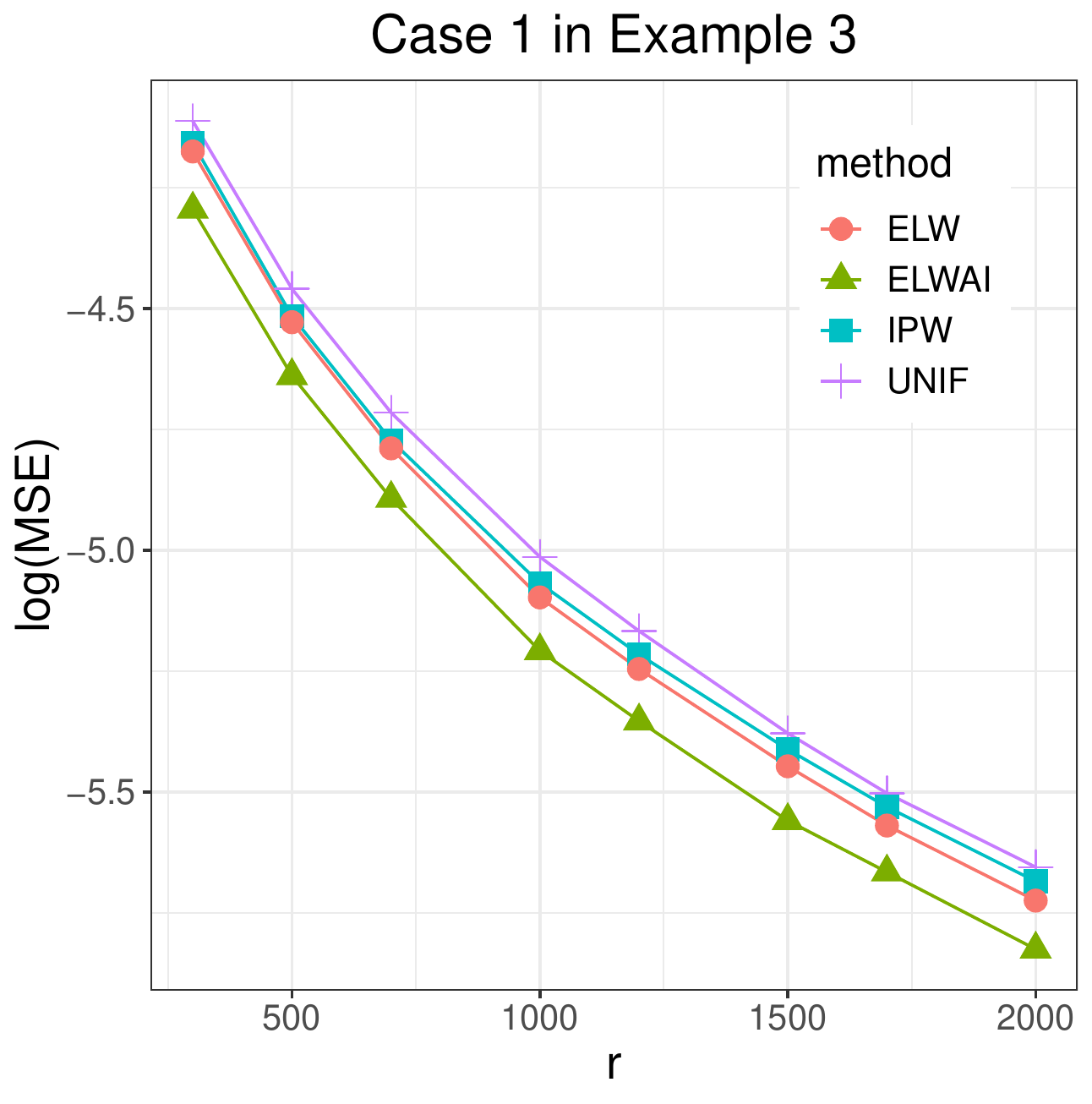}
\includegraphics[width= 3.7cm]{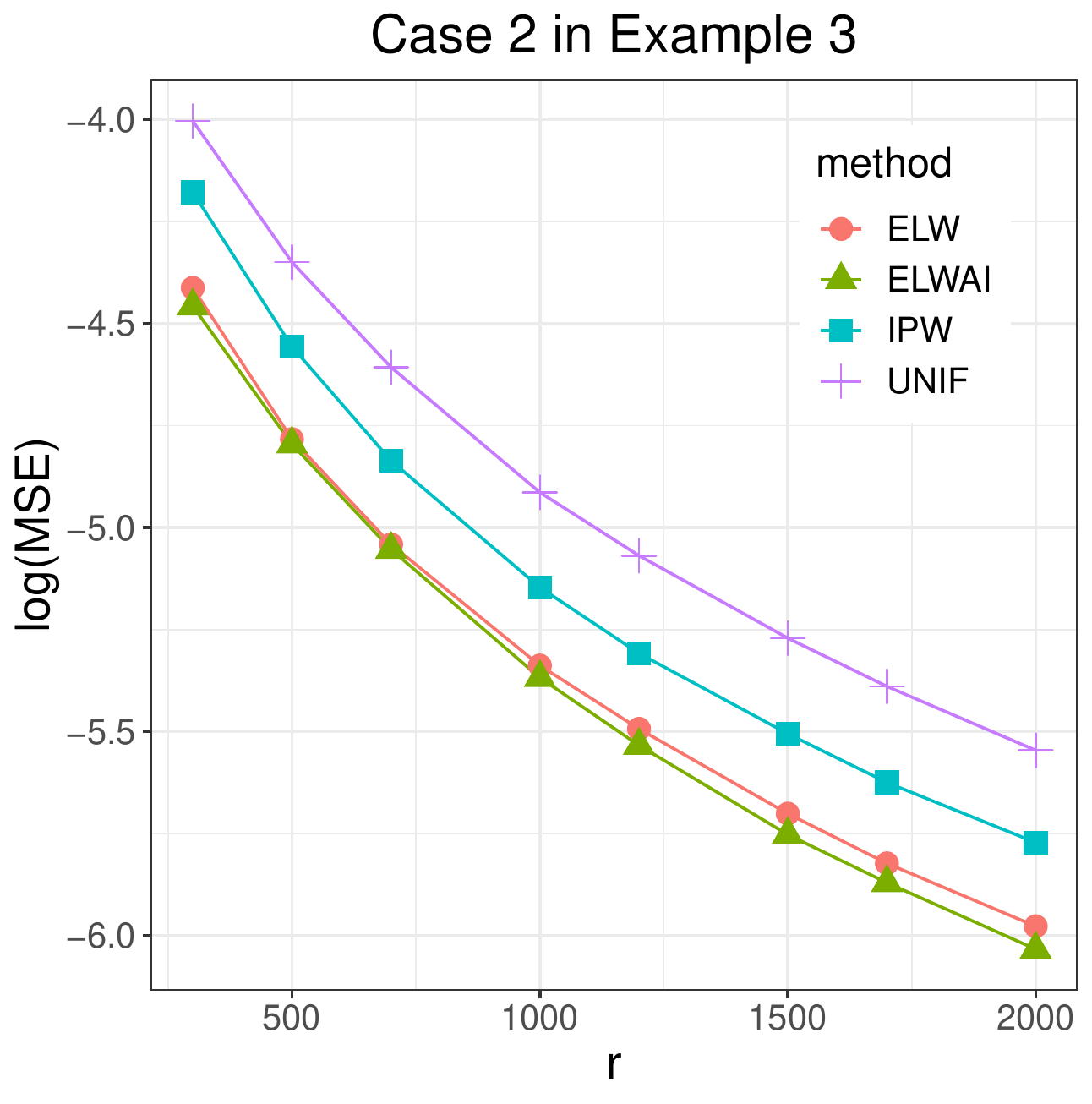}
\includegraphics[width= 3.7cm]{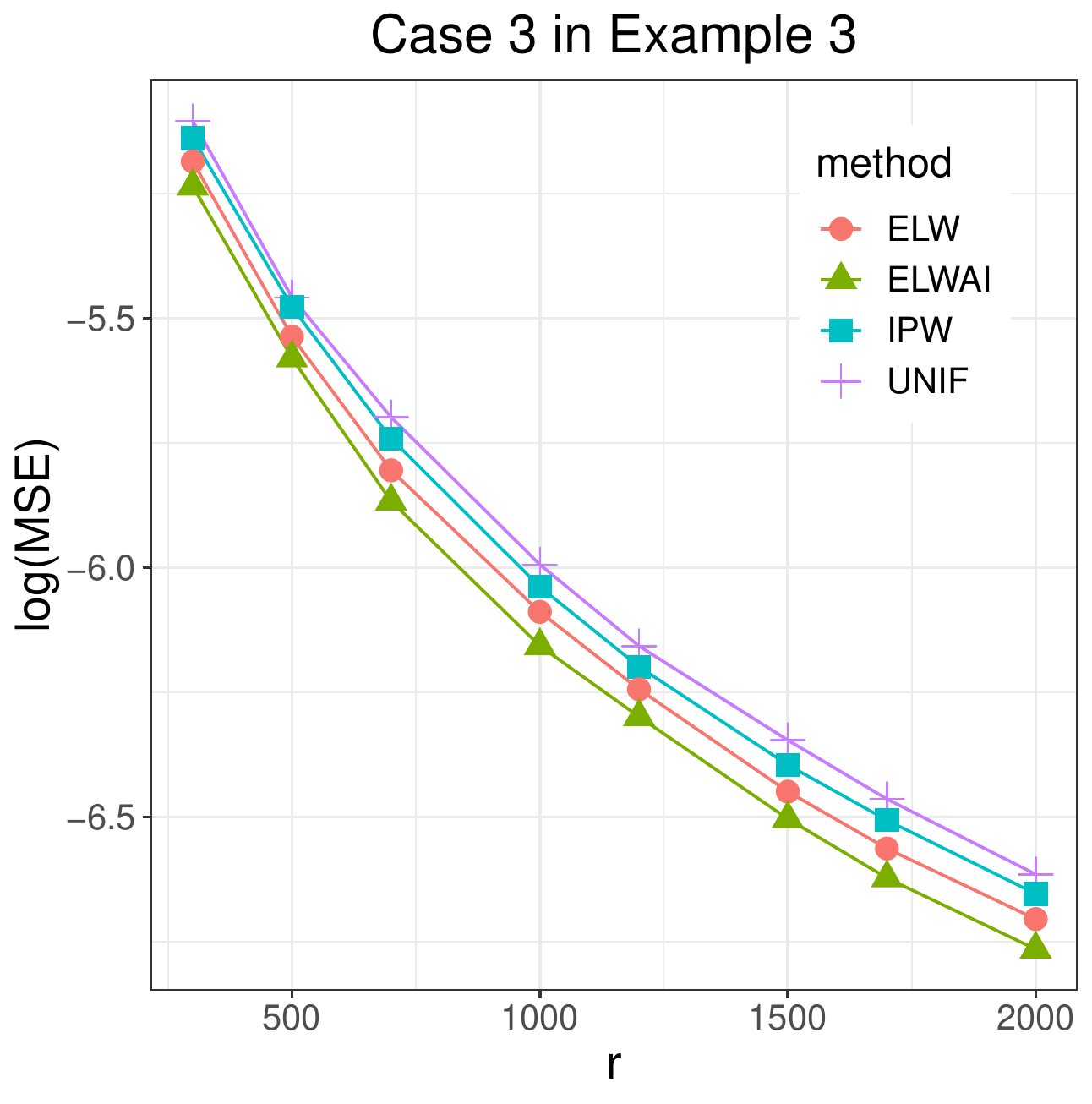}
\includegraphics[width= 3.7cm]{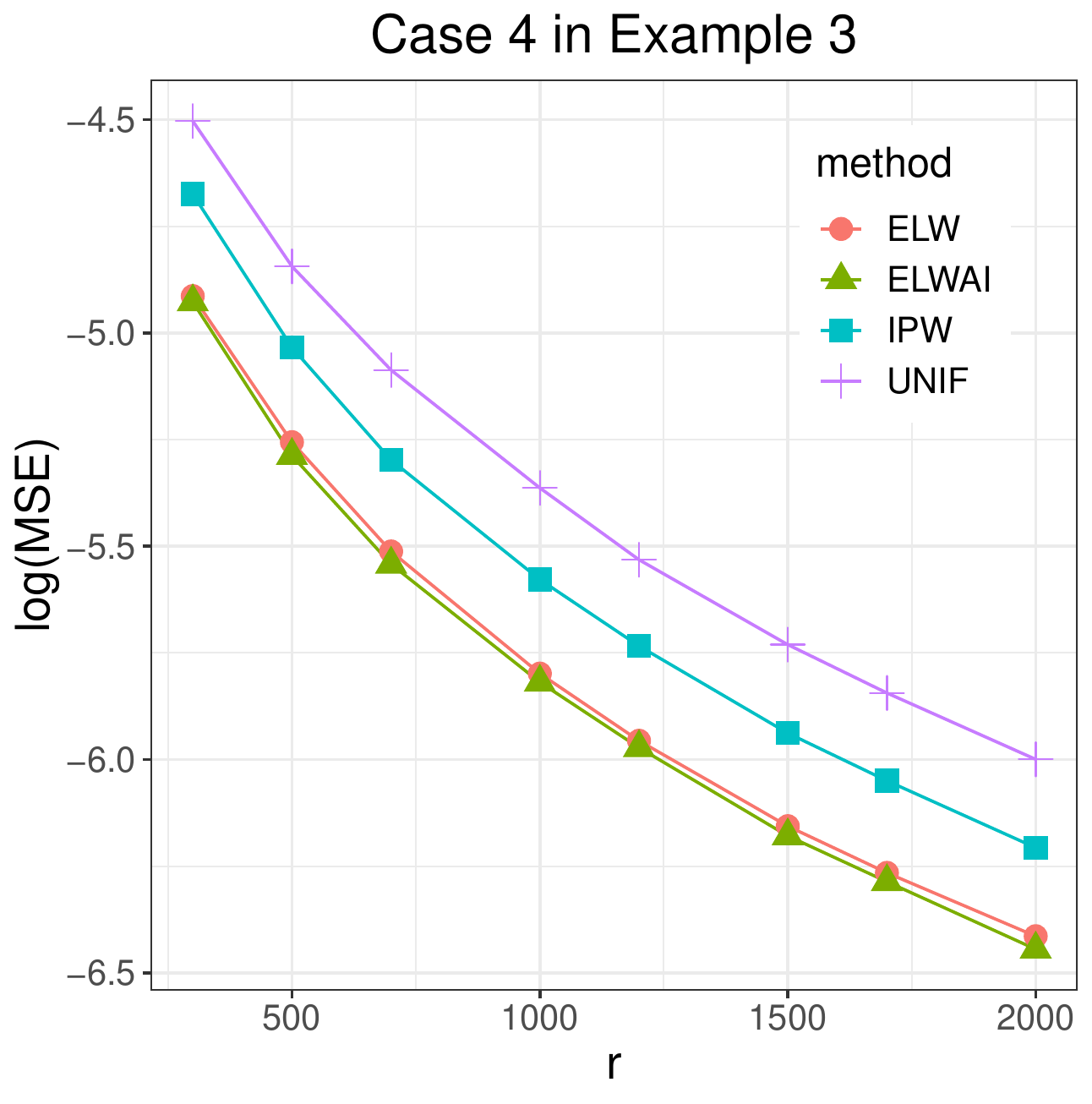}
\caption{Plots of the logarithm of MSE versus $r$ for
UNIF, IPW, ELW, and ELWAI under the L-criterion.}
\label{fig:sim-MSE-L}
\end{figure}

Under the A- and L-criteria, we generate 5000 subsamples
by each of the methods under comparison in each scenario of Examples 1--3.
We compare the performance of the methods
in terms of the empirical MSE
\ba\label{eq:mse-sim}
{\rm MSE} = \frac{1}{5000} \sum_{b=1}^{5000} \|\breve \theta_b - \hat\theta_N\|^2,
\ea
where $\breve \theta_b$ is a generic estimate
in the $b$th repetition and
$\hat\theta_N$ is the M-estimator based on the big data.
Figures \ref{fig:sim-MSE-A} and \ref{fig:sim-MSE-L}
display the logarithms of empirical MSE versus
$r$ under the A- and L-criteria, respectively.

We first examine the results in Figure \ref{fig:sim-MSE-A} under the A-criterion.
In this case, the empirical MSEs in \eqref{eq:mse-sim} are good approximations for
the asymptotic MSEs of the four estimators of $\theta$.
We take UNIF as the benchmark in handling big data,
because the uniform sampling involved does not reflect  any information about the big data.
Its most obvious advantage is requiring nearly no extra calculation cost.
An unequal probability sampling does not make sense
for big data analysis if
the resulting estimator is inferior to the UNIF-based estimator.
Figure \ref{fig:sim-MSE-A} shows that
ELW, ELWAI, and IPW all outperform UNIF
in terms of MSE uniformly for all $r$, indicating
that the ELW- and IPW-based two-step
unequal probability samplings seem to be meaningful.
Moreover, both ELW and ELWAI outperform IPW uniformly for all $r$,
although a shrinkage technique is employed
for IPW \citep{ma2014statistical} in Examples 1 and 2.
This suggests that the proposed ELW estimation and
nearly optimal sampling strategy
produce better estimators than
the IPW estimation and sampling strategy,
regardless of whether auxiliary information is used.
In particular, the estimation efficiency gains of ELW and ELWAI over IPW
are remarkable,
except in cases 1 and 3 of Example 3.
Regarding the two ELW methods, the ELWAI-based estimator gives a uniformly smaller MSE
than the ELW-based estimator, especially in case 1 of Example 3.
This clearly implies the ELW method can produce more reliable estimators
by incorporating auxiliary information, as disclosed by Theorem \ref{asym-el}.

When the A-criterion is replaced by the L-criterion,
the empirical MSEs in \eqref{eq:mse-sim} of the generic estimator $\breve \theta$
are different from the asymptotic MSEs of the linearly transformed estimator $V \breve \theta$.
The optimal sampling plan minimizing the latter may
not produce a point estimator that has a minimal empirical MSE.
Even so, the results in Figure \ref{fig:sim-MSE-L} show that
the efficiency order of ELW, ELWAI, IPW, and UNIF
is the same as in Figure \ref{fig:sim-MSE-A},
indicating that the proposed ELW methods uniformly outperform IPW again.
Additionally, by incorporating auxiliary information, ELWAI
achieves an efficiency gain over ELW.
One benefit of using the L-criterion is that
ELW, ELWAI, and IPW have much lower computational costs than
under the A-criterion.

\subsection{Evaluation of our sample size determination methods}
\label{sec:sim-size}

In Section 4, we presented two sample size determination methods, M1 and M2,
under requirements (R1) and (R2), respectively.
With the sample sizes determined by M1 and M2,
we now examine whether the proposed sampling and estimation strategy produces
estimators that have the desired precision.
To this end, we fix the first-capture sample size to $r_0 = 200$ and
determine the second-capture sample size $r$ by
$\tilde r = N(\tilde n_0 - r_0)/(N - r_0)$ for the ELW method,
where $\tilde n_0$ is the root of
\eqref{eq:n_0} under requirement (R1) or \eqref{eq:n_0_accu}
under requirement (R2), where $a=5\%$.

\begin{figure}[h]
\centering
\includegraphics[width= 4.8cm]{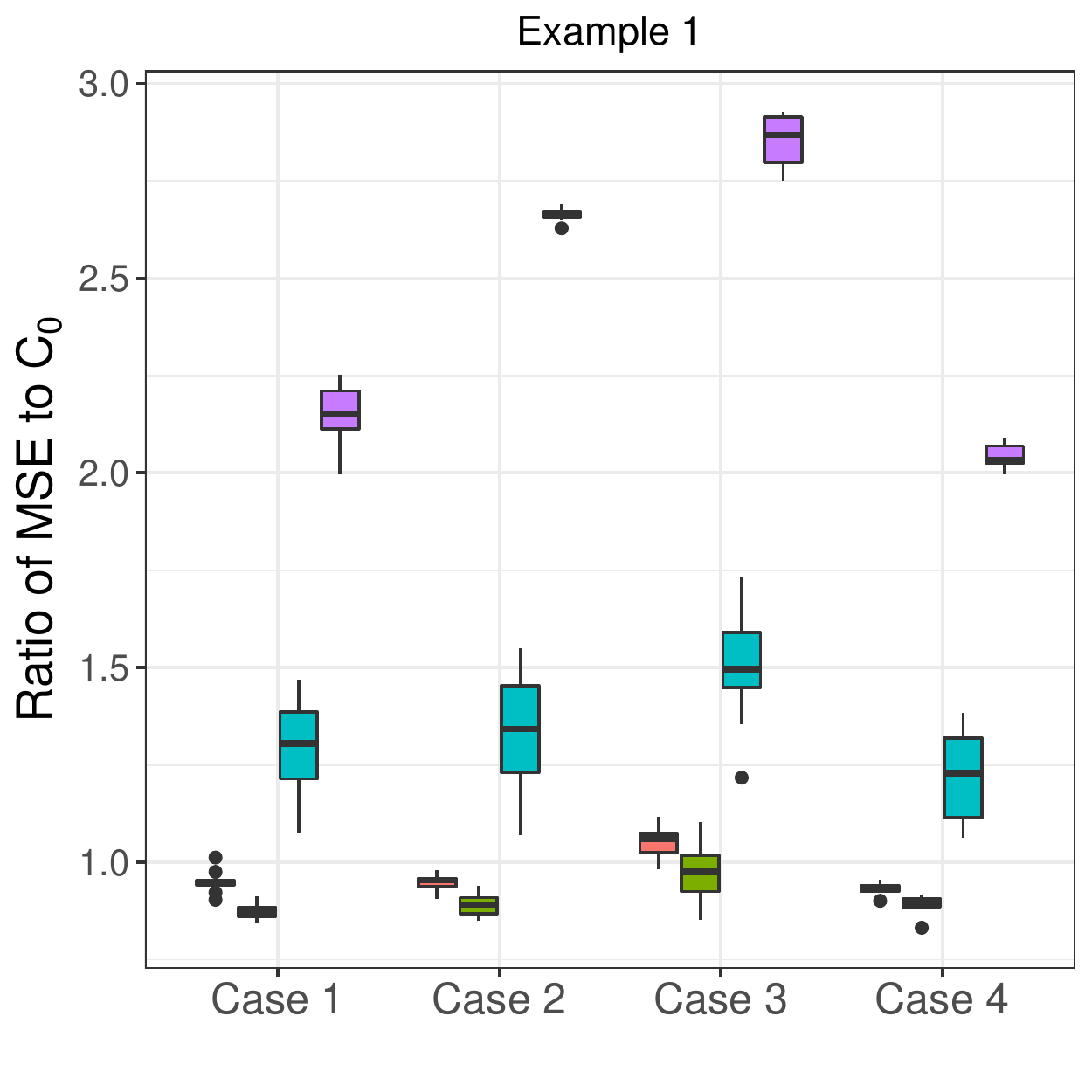}\hspace{.5em}
\includegraphics[width= 4.8cm]{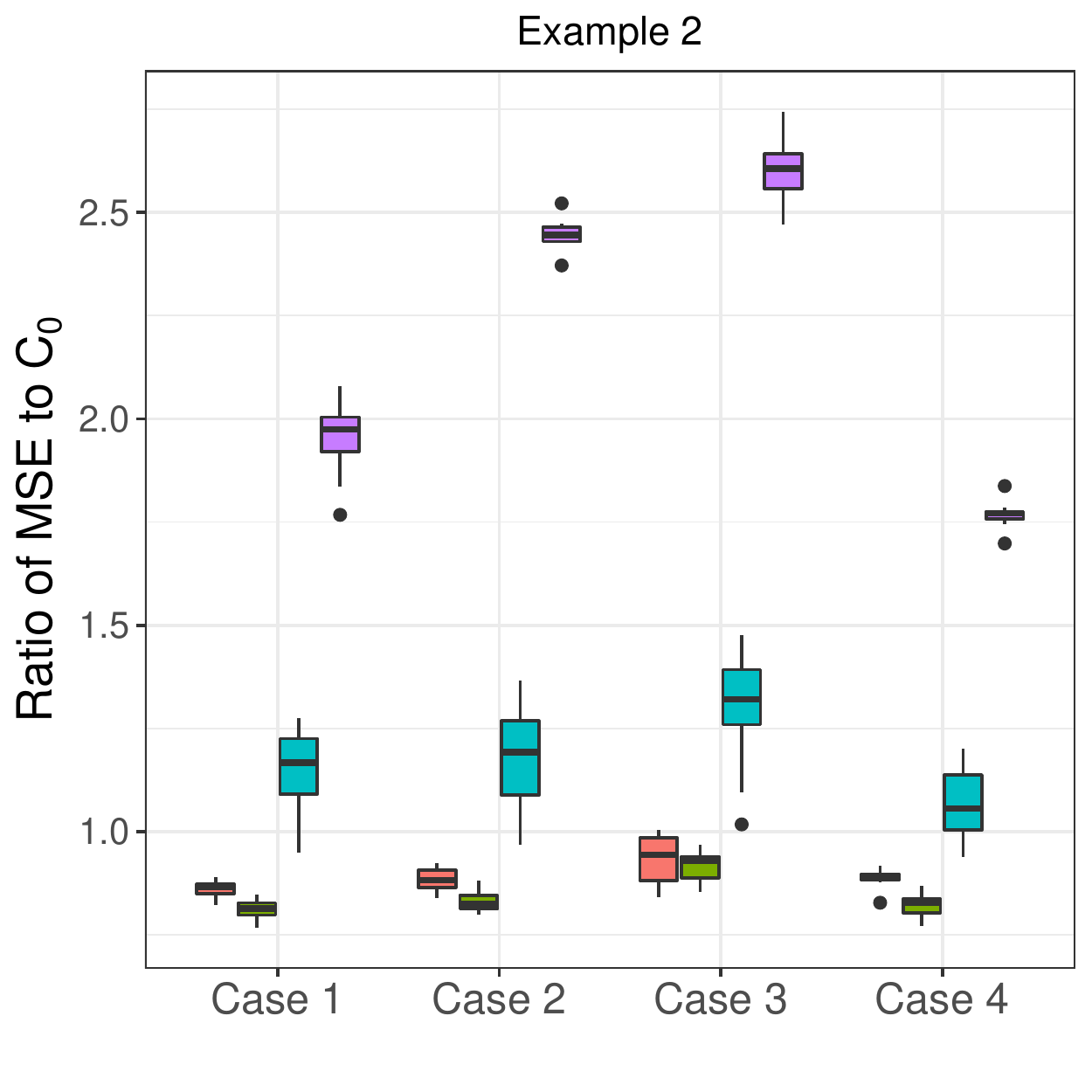}\hspace{.5em}
\includegraphics[width= 4.8cm]{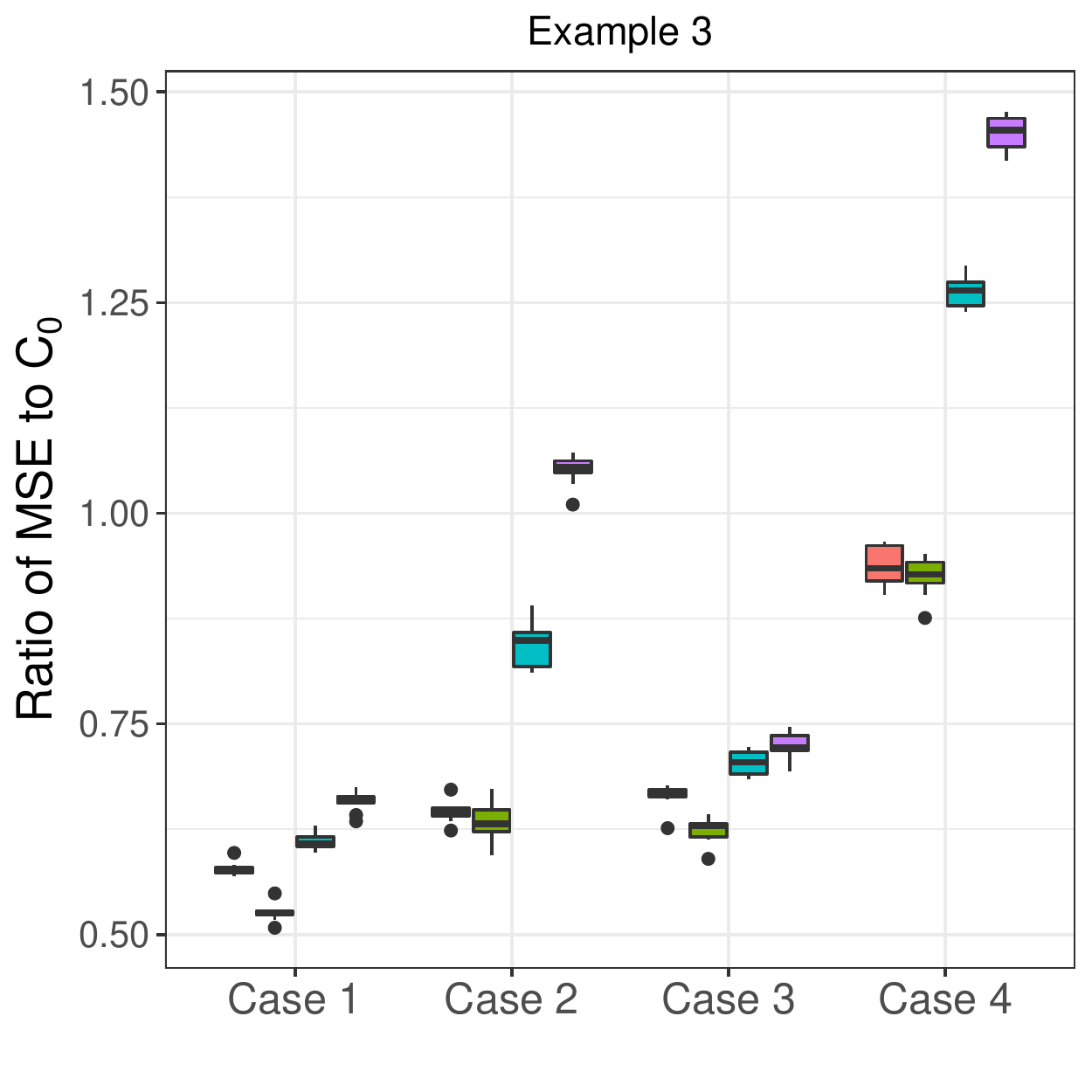}\\
\includegraphics[width= 4.8cm]{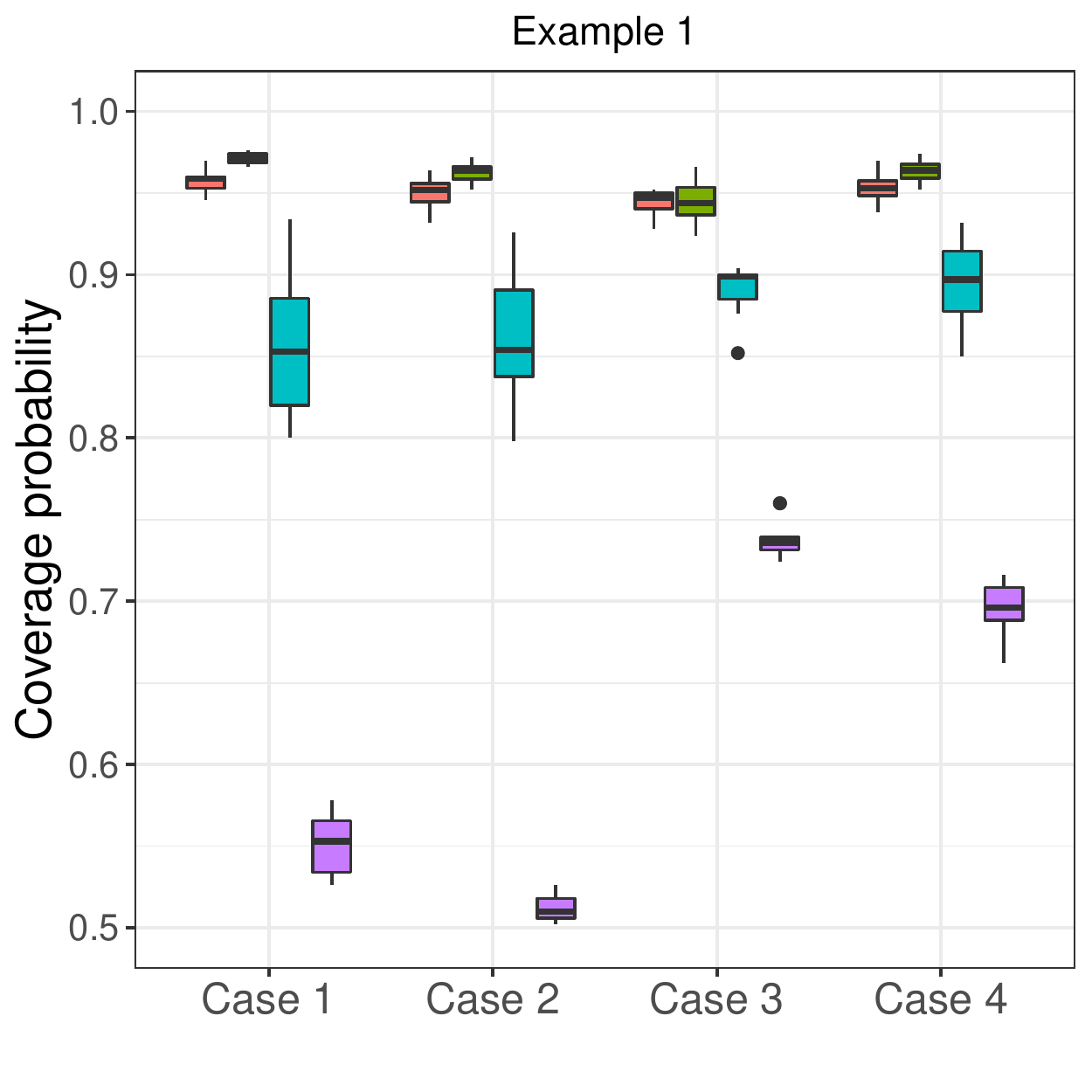}\hspace{.5em}
\includegraphics[width= 4.8cm]{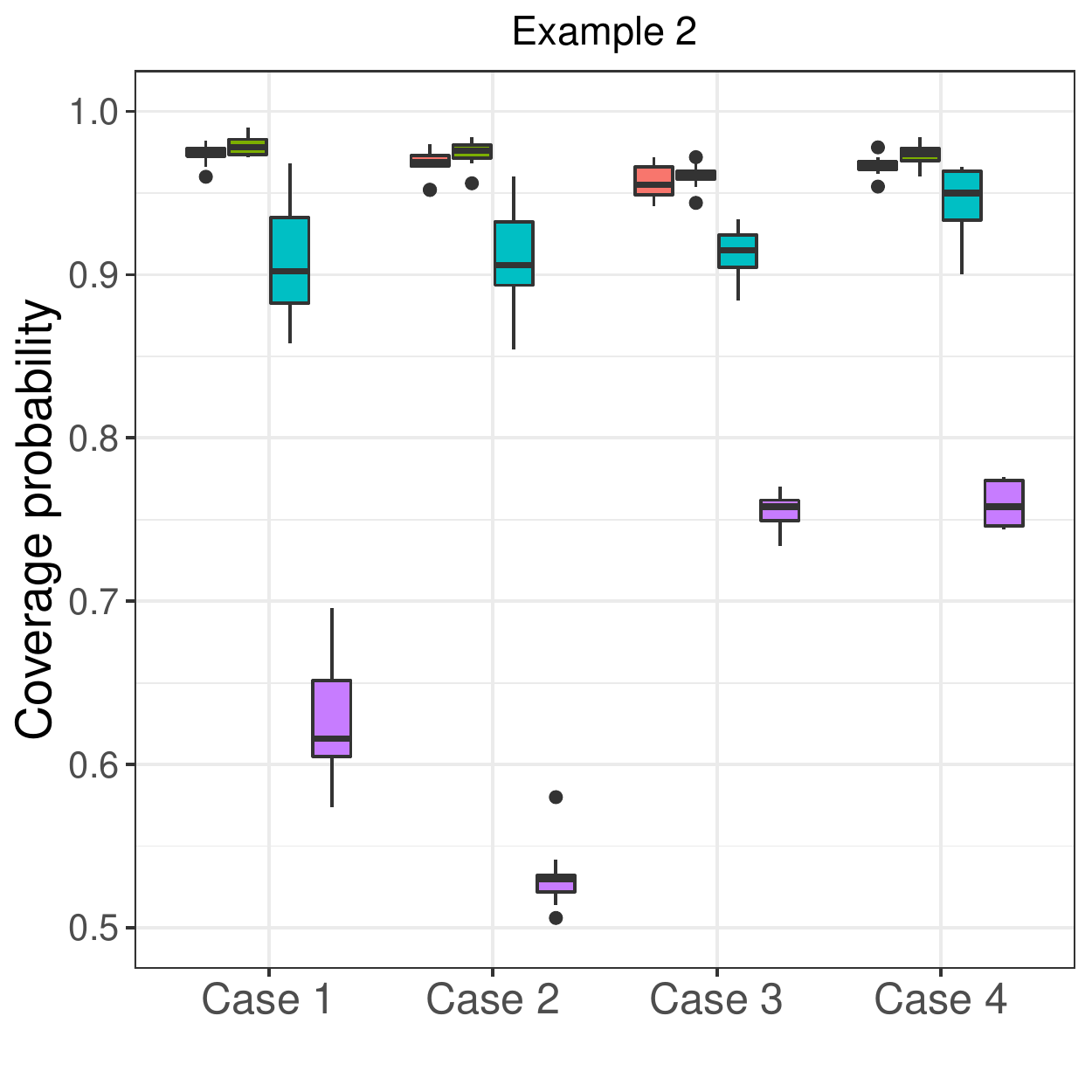}\hspace{.5em}
\includegraphics[width= 4.8cm]{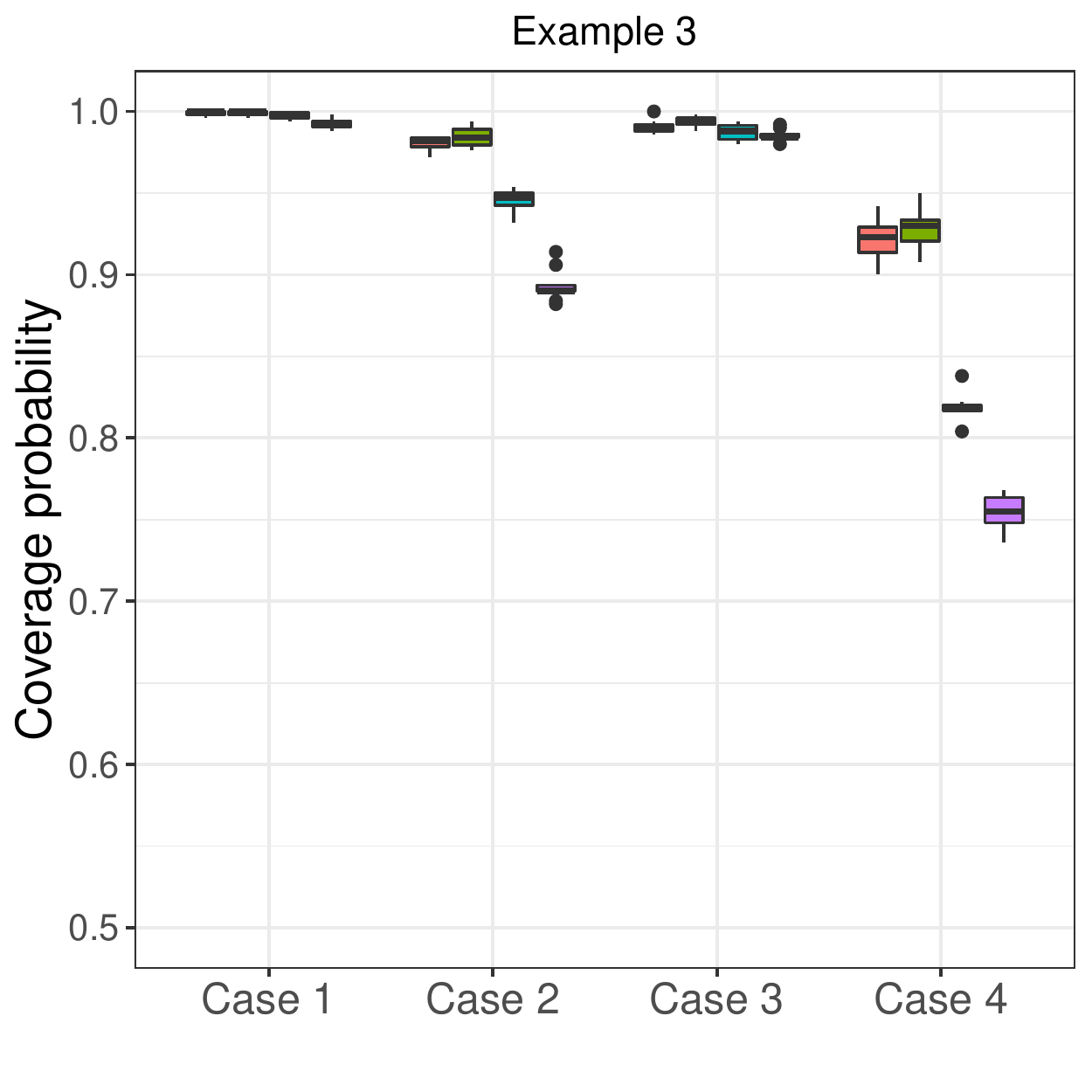}
\caption{
Ratios of actual MSEs to $C_0$ under requirement (R1) (upper panel)
and simulated coverage probabilities under requirement (R2) (lower panel)
with $1-a = 95\%$.
ELW (red), ELWAI (green), IPW (blue),
and UNIF (purple)
correspond to columns 1--4 from left to right in each case. }
\label{fig:sim-size}
\end{figure}

We consider 10 distinct values
of $C_0$ or $d_0$ for each case,
so that $\tilde r$ ranges from 300 to 2000.
To ensure a fair comparison, we apply ELWAI and IPW
with same sample size pair $(r_0, \tilde r)$,
namely the ideal size of the initial sample is $r_0$
and that for the second sample is $\tilde r$.
When applying UNIF, we set the total sample size to be $r_0+\tilde r$.
For a generic estimator $\breve \theta$,
we calculate the ratio of its actual MSE to
the specified $C_0$,
and simulate the coverage probabilities of
$\{\theta: \|\breve\theta - \theta\|\leq d_0\}$ 
based on 500 simulated repetitions.
The results are displayed in
Figure \ref{fig:sim-size},
where each box-plot is based on 10 ratios (upper panel)
or simulated coverage probabilities (lower panel).

In the upper panel, the ratios based on ELW are all close to or less than 1.
In other words, with the sample size determined by M1,
the MSE of the ELW-based estimator is close to (and
no greater than) the prespecified precision $C_0$.
The sample size is quite accurate
under Examples 1 and 2 because
the ratios are quite close to 1,
although it is somewhat conservative
in cases 1--3 of Example 3, where the ratios are no greater than 75\%.
The ratios based on ELWAI are always slightly smaller than those based on ELW,
which makes sense as the ELWAI-based estimator is more
efficient than the ELW-based estimator, both theoretically and numerically.
The ratios based on IPW and UNIF are much greater than 1,
which coincides with the observation that they both are less efficient than ELW.

In the lower panel, the coverage probabilities corresponding to ELW
are always close to or greater than 95\%, as specified by requirement (R2).
Thus, with the sample size determined by M2,
the confidence region $\{\theta: \|\hat\theta_{\elw} - \theta\|\leq d_0\}$ 
has coverage probabilities
no less than the prespecified confidence level of $95\%$.
With the same sample size, the ELWAI-based confidence region has an even
greater coverage probability, because the ELWAI-based estimator is more
efficient than the ELW-based estimator.
However, the IPW- and UNIF-based confidence regions have much low coverage probabilities,
which are often no greater than 90\%.
In all cases, the boxplots of ELW and ELWAI are
much shorter than those of IPW and UNIF,
suggesting that ELW and ELWAI provide
much more stable performance than IPW and UNIF.

Overall, M1 and M2 usually produce reasonable sample sizes
with which the ELW method approximately meets target
requirements (R1) and (R2), respectively.
The inferior performance of IPW and UNIF in Figure \ref{fig:sim-size} indicates that
many more samples
are usually required to
achieve the same estimation precision compared with ELW and ELWAI.

\section{Applications}\label{sec:data}
In this section, we further investigate the performance
of the proposed ELW estimation and nearly optimal capture--recapture
sampling method by analyzing three real datasets:
a bike sharing dataset, an income dataset,
and a protein structure dataset, as found in the supplementary material of \cite{yao2021review}.

The bike sharing dataset consists of
17,379 observations,
in which we take
the number of bikes rented hourly as the response.
The covariates include
a binary variable $X_1$,
indicating whether a certain day is a working day or not,
and the three continuous variables
of temperature ($X_2$), humidity ($X_3$), and windspeed ($X_4$).
The income dataset contains
48,842 observations, in which
the response is
a binary variable indicating
whether one person's income is over 50,000 \$ or not.
Five continuous covariates are considered, namely
the person's age, weight, education, capital loss, and working hours per week,
denoted as $X_1$--$X_5$, respectively.
The protein structure dataset contains
45,730 observations, where the 75-th percentile of
the size of the residue ranging
from 0 to 21 Angstrom may be affected by eight covariates ($F_1$, $F_2$, $F_4$--$F_9$),
denoted as $X_1$--$X_8$.

To investigate the relationship between the responses and the covariates,
we fit a Poisson regression model,
a logistic regression model, and
a quantile regression model with $\tau = 0.75$
to the three datasets, respectively.
To eliminate the influence of scales of different variables,
we centralize and standardize the covariates in all datasets
and the response variable in the protein structure dataset.
The regression coefficients of the regression models
based on the full datasets are reported in Table \ref{tab:dat-full}.

\begin{table*}[h]
\centering \scriptsize
\setlength{\columnsep}{.05in}
\caption{Full-data-based estimates of regression coefficients.}
\label{tab:dat-full}
\begin{threeparttable}
\setlength{\tabcolsep}{5.5pt}{
\begin{tabular}{cccrrrrrrrr}
\toprule
Data &Model& Intercept & $X_1$& $X_2$& $X_3$& $X_4$
& $X_5$& $X_6$& $X_7$& $X_8$
\\
\midrule
Bike sharing &Poisson regression&
5.02& 0.03& 1.83& -1.36& 0.20
\\
Income &Logistic regression&-8.59&0.05&6E-7 
&0.34&6E-4& 0.04
\\
Protein structure & Quantile regression&
0.62& 0.89 & 0.87& -1.27& -0.38& -0.38& -0.04& 0.26& -0.10
\\
\bottomrule
\end{tabular}}
\end{threeparttable}
\end{table*}

\begin{figure}[h]
\centering
\includegraphics[width= 4.7cm]{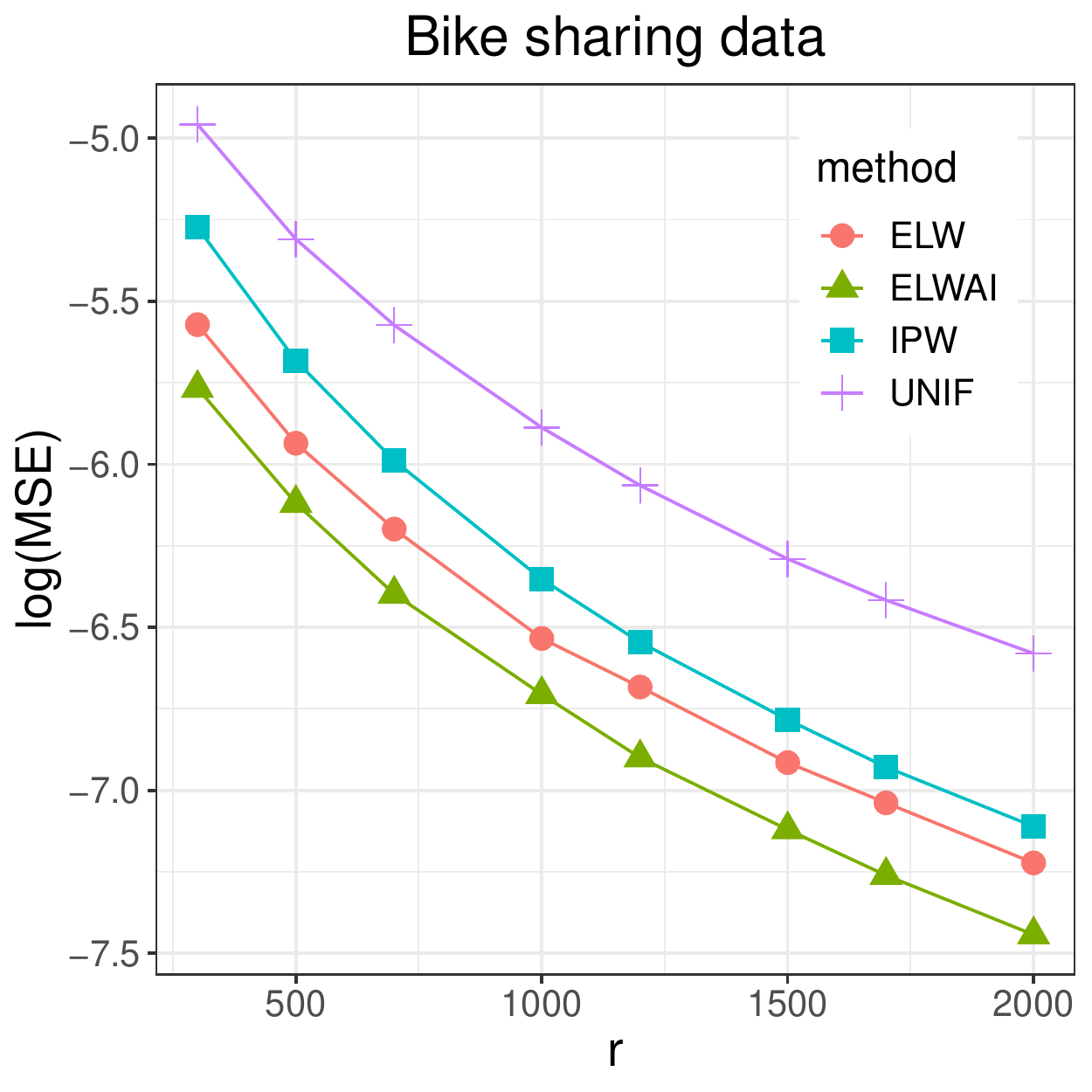}
\includegraphics[width= 4.7cm]{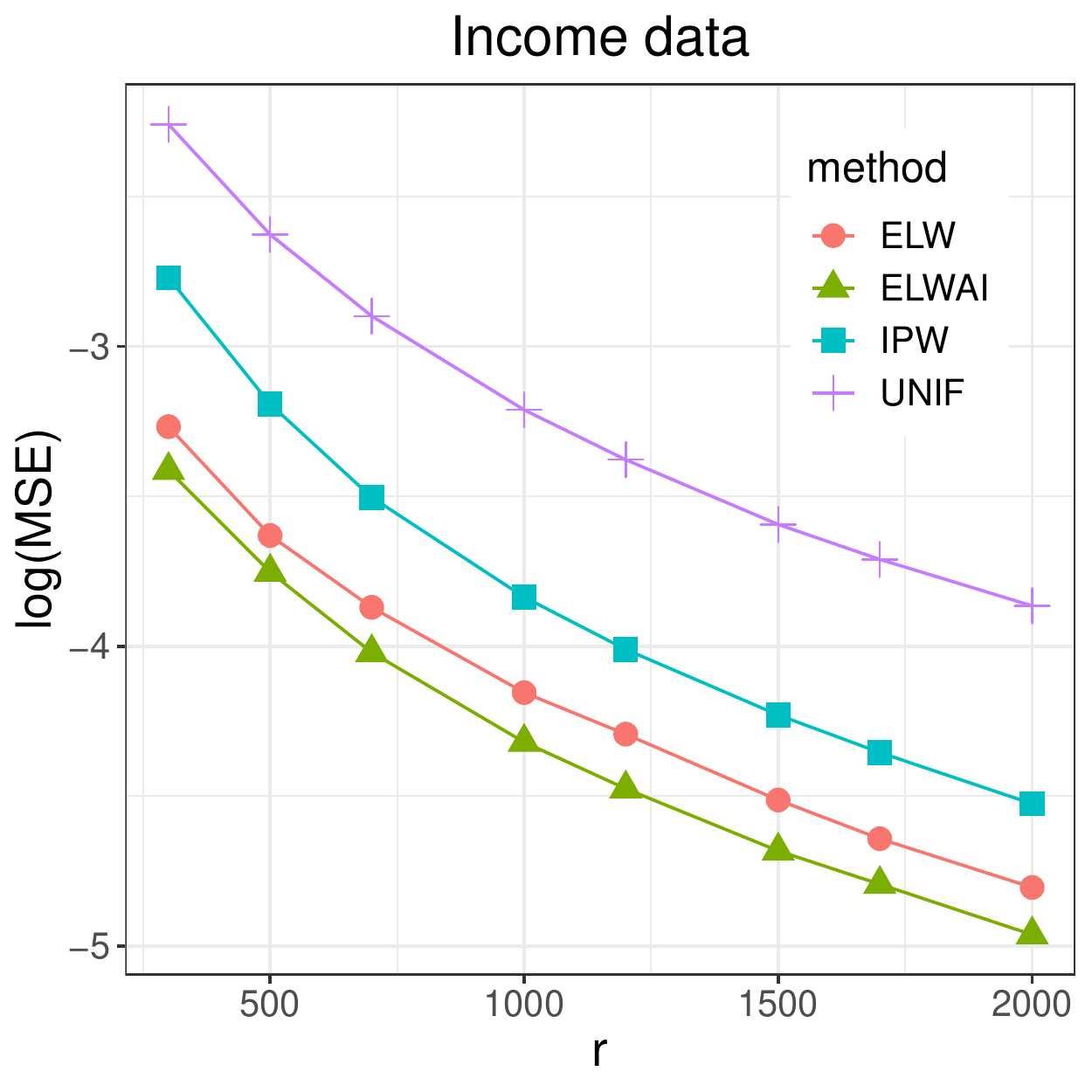}
\includegraphics[width= 4.7cm]{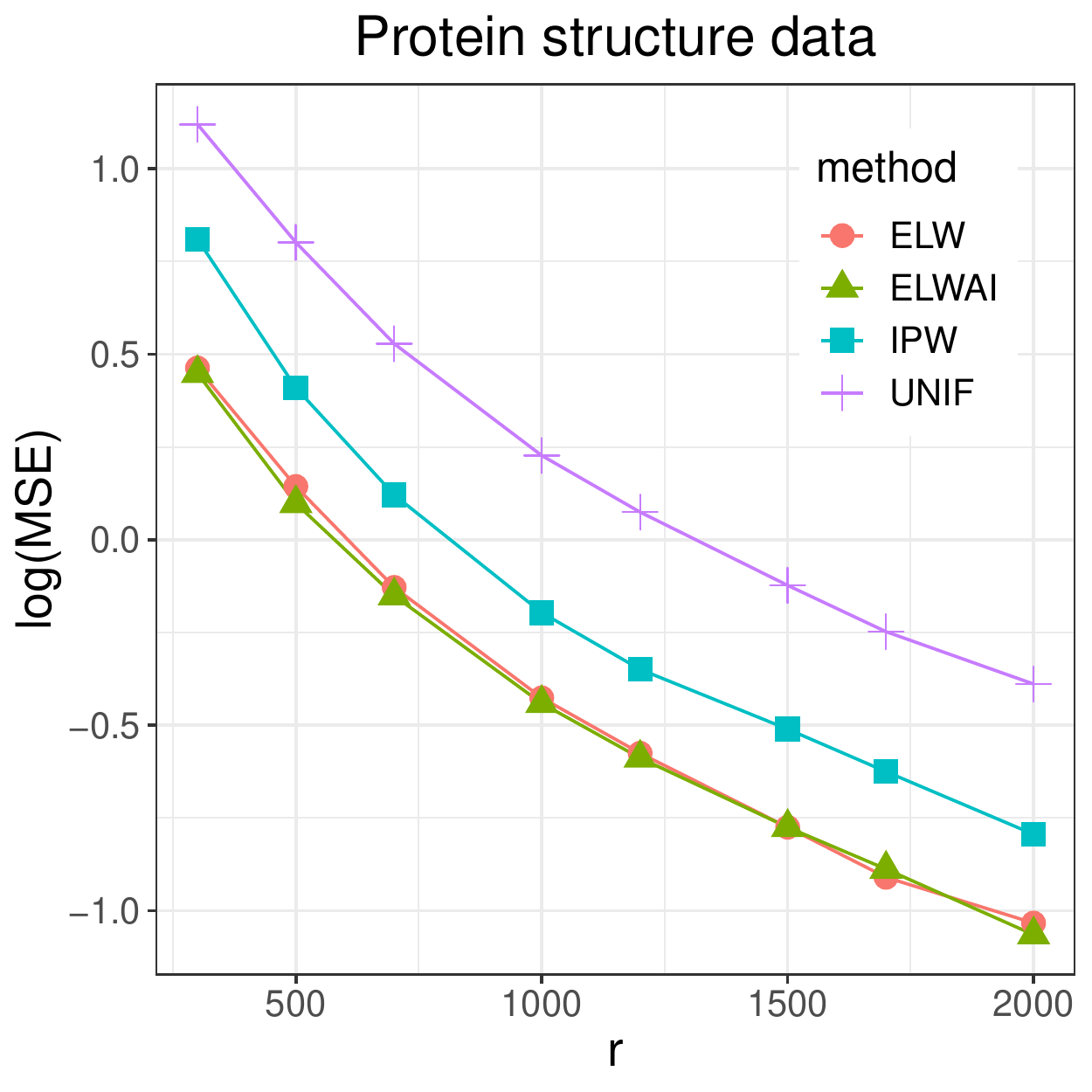}
\\
\includegraphics[width= 4.7cm]{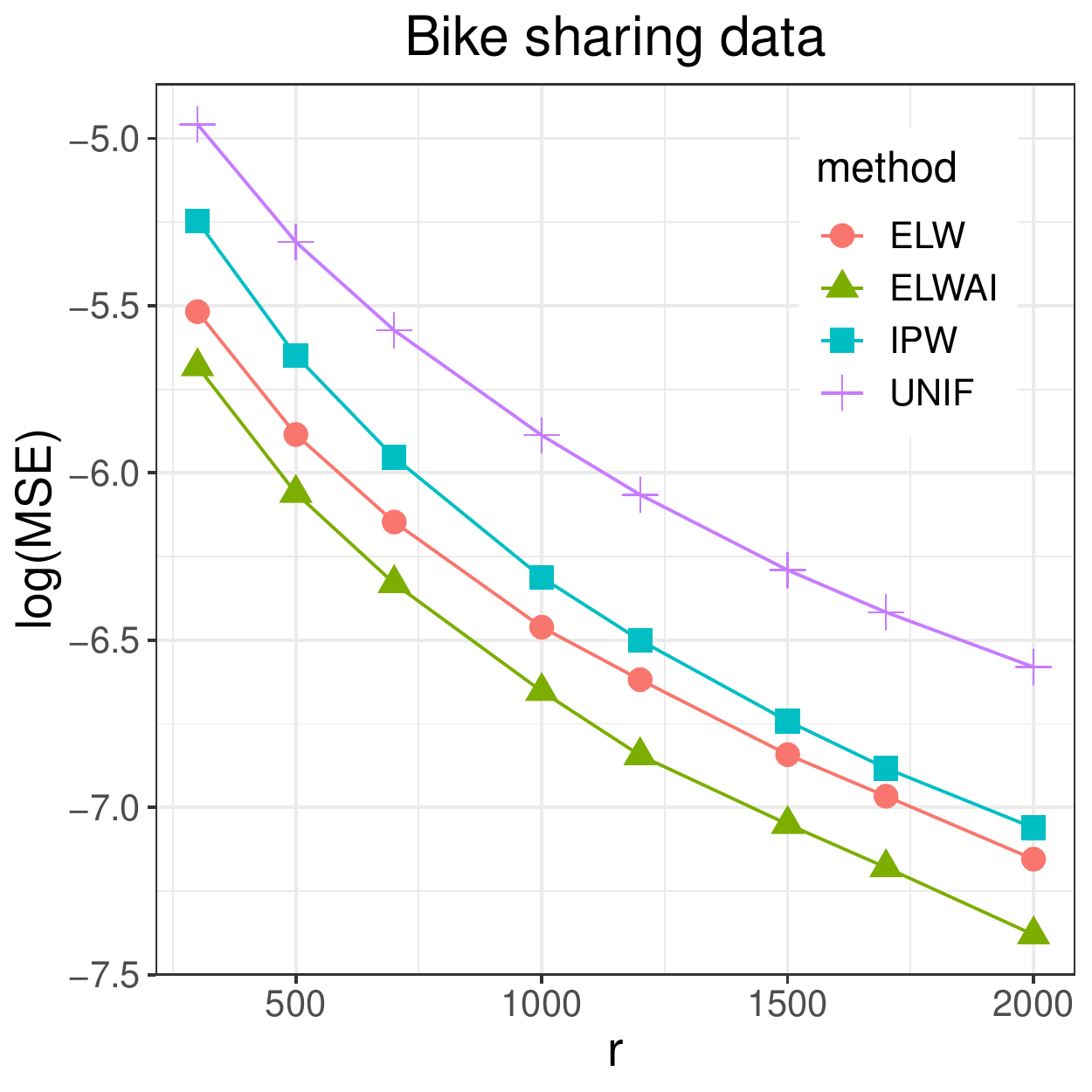}
\includegraphics[width= 4.7cm]{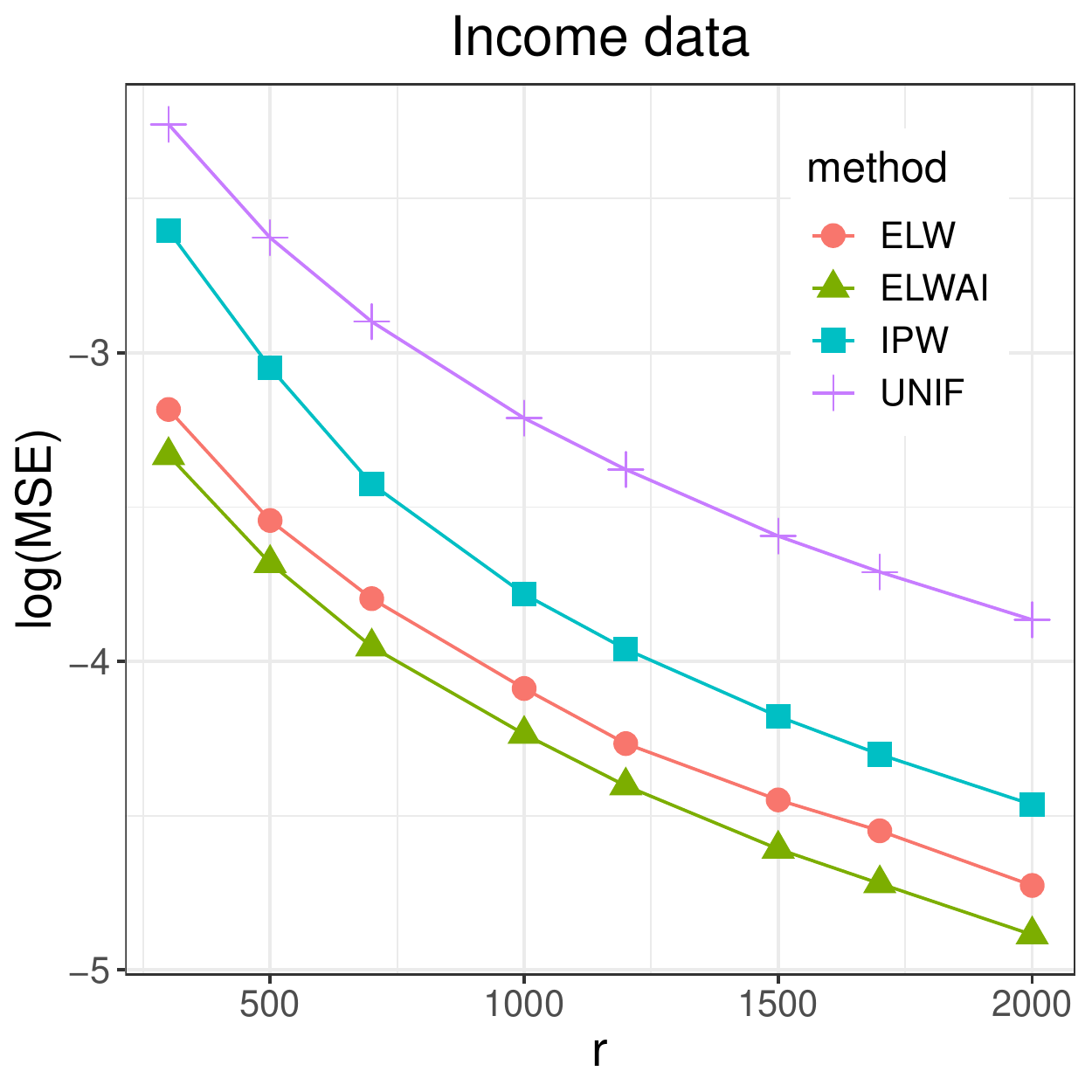}
\includegraphics[width= 4.7cm]{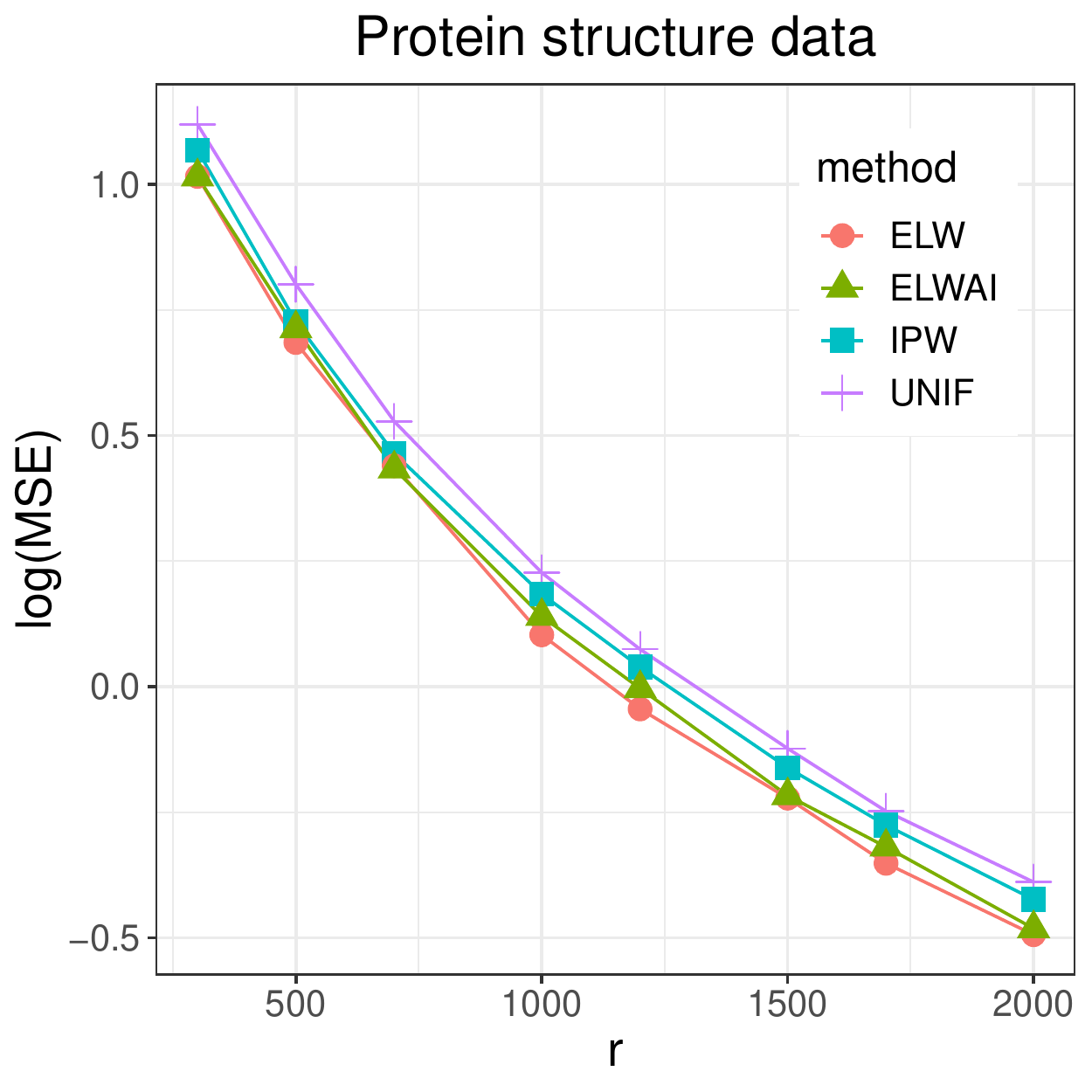}
\caption{
Plots of the logarithm of MSE versus $r$ for
UNIF, IPW, ELW, and ELWAI under the A-criterion (upper row) and the L-criterion (lower row)
based on the three real datasets. }
\label{fig:sim-MSE}
\end{figure}

We apply the UNIF, IPW, ELW, and ELWAI methods to the three real datasets.
The remaining settings, such as sample sizes and number of simulation repetitions,
are the same as those in Section \ref{eq:sim-eff}.
Figure \ref{fig:sim-MSE}
displays the logarithms of the empirical MSEs of a point estimator
versus the sample size $r$ of the second-step sampling.
Clearly, ELW, ELWAI, and IPW
outperform the naive method, UNIF, by a large margin
when the A-criterion is used to construct the optimal sampling design.
The proposed ELW- and ELWAI-based estimators both have much smaller empirical MSEs.
These findings can also be seen from
the results for the bike sharing dataset and the income dataset under the L-criterion,
and coincide with those from our simulation studies.
What differs is that the efficiency gains of ELWAI over ELW
based on the first two datasets are
much greater than those in our simulation studies.
This implies that the auxiliary information of the full-data response
contains more information, and is thus more helpful in improving the performance of ELW
in the former than in the latter.
Based on the protein structure dataset under the L-criterion,
although all four methods have almost the same performance,
our ELW and ELWAI are slightly more reliable than IPW and UNIF.

Based on the three real datasets, we also investigate
the performance of our sample size determination methods, M1 and M2,
under the same settings as in Section 5.3.
Again we find that  they
provide desirable sample sizes that guarantee
the ELW and ELWAI methods meet the given estimation precision requirements.

\section{Discussion}\label{sec:disc}

Based on a capture--recapture sample from a big dataset,
we have developed an ELW estimation method for M-estimation problems.
The proposed approach not only overcomes the instability of the conventional IPW
estimation method, but also improves the estimation efficiency
by incorporating auxiliary information.
A nearly optimal capture--recapture sampling plan was constructed accordingly.
Theoretically, the ELW method is asymptotically more efficient than the IPW method,
which means that the proposed sampling and estimation method
requires fewer samples to achieve the target estimation precision.
For technical convenience, we assumed the convexity of the loss function in the M-estimation problem.
Our ELW estimation method also applies to
general M-estimation problems and general estimating equation problems
\citep{qin1994empirical}.
Further efforts may be needed to establish the asymptotic normality
of the resulting point estimator,
which is the foundation for constructing optimal sampling plans.

The capture--recapture sampling we have considered consists of
a pilot uniform sampling and a refined sampling.
Under this sampling framework,
we established two sample size determination methods
under estimation precision requirements (R1) and (R2), respectively.
These methods are new in the literature of optimal subsampling for big data.
They may need to be modified when the parameter of interest
is a smooth function of $\theta$, such as $C \theta$ for a given matrix $C$,
rather than $\theta$ itself.
In addition, the current capture--recapture sampling consists of only two subsampling processes,
although this may be extended to multiple subsampling processes when needed.

\bigskip
\begin{center}
{\large\bf SUPPLEMENTARY MATERIAL}
\end{center}

\begin{description}

\item[]
The supplementary material
contains the proofs of Theorems \ref{asym-ipw}--\ref{asym-el-cap*}. (SubsampEL$\_$supp.pdf)

\end{description}

\bibliographystyle{natbib}

\end{document}